\documentclass[twocolumn,floatfix,superscriptaddress,a4paper,showkeys,nofootinbib,reprint]{revtex4-1}
\usepackage{epsfig}
\usepackage{latexsym}
\usepackage{xspace}
\usepackage[colorlinks=true,linktocpage=true,linkcolor=blue,citecolor=blue,allcolors=blue]{hyperref}
\usepackage[utf8]{inputenc}
\usepackage{indentfirst}
\usepackage{enumerate}
\usepackage{listings}
\usepackage{color}

\usepackage[caption=false,position=top]{subfig}

\usepackage{amsmath}
\usepackage{amssymb}
\usepackage[english]{babel}
\usepackage{url}

\newcommand{\eq}[1]{\begin{align} #1 \end{align}}
\newcommand{\mean}[1]{\langle #1 \rangle}

\newcommand{\sNN}{\sqrt{s_{\rm NN}}}

\usepackage[normalem]{ulem}  % \sout{old text} for strikeout

\begin{document}

\title{High-order cumulants and correlation functions\\ near the critical point from molecular dynamics}

\author{Volodymyr~A.~Kuznietsov}
    \affiliation{Physics Department, University of Houston, 3507 Cullen Blvd, Houston, TX 77204, USA}
    \affiliation{Department of the high-density energy physics, Bogolyubov Institute for Theoretical Physics, 03680 Kyiv, Ukraine}

\author{Roman Poberezhniuk}
    \affiliation{Physics Department, University of Houston, 3507 Cullen Blvd, Houston, TX 77204, USA}
    \affiliation{Department of the high-density energy physics, Bogolyubov Institute for Theoretical Physics, 03680 Kyiv, Ukraine}

\author{Mark~I.~Gorenstein}
    \affiliation{Department of the high-density energy physics, Bogolyubov Institute for Theoretical Physics, 03680 Kyiv, Ukraine}
    \affiliation{Extreme Matter Institute EMMI, GSI Helmholtzzentrum für Schwerionenforschung GmbH, Planckstrasse 1, 64291 Darmstadt, Germany}
    
\author{Volker Koch}
\affiliation{Nuclear Science Division, Lawrence Berkeley National Laboratory, 1 Cyclotron Road, Berkeley, CA 94720, USA}

\author{Volodymyr~Vovchenko}
\affiliation{Physics Department, University of Houston, 3507 Cullen Blvd, Houston, TX 77204, USA}

\begin{abstract}
We present a systematic investigation of particle number fluctuations in the crossover region near the critical endpoint of a first-order phase transition using molecular dynamics simulations of the classical Lennard–Jones fluid.
We extend our prior studies to third- and fourth-order cumulants in both coordinate- and momentum-space acceptances and integrated correlation functions (factorial cumulants).
We find that, even near the critical point, non-Gaussian cumulants equilibrate on time scales comparable to those of the second-order cumulants, but show stronger finite-size effects.
The presence of interactions and of the critical point leads to strong deviations of the cumulants from the ideal-gas baseline in coordinate space; these deviations are expected to persist in momentum space in the presence of collective expansion.
In particular, the kurtosis becomes strongly negative, $\kappa \sigma^2 \ll -1$, on the crossover side of the critical point.
However, this signal is significantly diluted once an efficiency cut used to distinguish protons from baryons is applied, leading to $|\kappa \sigma^2| \lesssim 1$ even in the presence of the critical point.
We discuss our results in the context of ongoing measurements of proton number cumulants in heavy-ion collisions in RHIC-BES-II.
\end{abstract}


\maketitle

\section{Introduction}
\label{sec-intro}
A critical point (CP) marks the termination of a first-order phase transition line, where distinct phase boundaries disappear \cite{Landau1996}. This phenomenon is widespread across various physical systems, including atomic and molecular substances, ferromagnets, cold nuclear matter, and possibly hot QCD matter. A key characteristic of the CP is the enhancement of thermal fluctuations in its vicinity, which diverge in an infinite system. For example, the pronounced density fluctuations near the CP of a liquid-gas transition give rise to the well-known effect of critical opalescence.

Understanding the existence and location of the QCD critical point at finite baryon density is a central objective of modern heavy-ion collision experiments, such as the Beam Energy Scan (BES) program at RHIC~\cite{Bzdak:2019pkr}. Event-by-event fluctuations of the proton number are among the key observables in this search~\cite{Stephanov:1999zu, Hatta:2003wn}, with higher-order cumulants expected to exhibit nonmonotonic behavior as a function of collision energy if the CP is within experimental reach~\cite{Stephanov:2008qz, Stephanov:2024xkn}. 
Measurements from RHIC-BES-I~\cite{STAR:2021fge} and data from RHIC-BES-II~\cite{STAR:2025zdq} indicate a possible nonmonotonic trend in fourth-order (factorial) cumulants of net-proton distributions~\cite{STAR:2021fge}, although statistical uncertainties remain significant. 
Additionally, the measurements of second-order cumulants have also revealed an excess in the scaled variance of proton number fluctuations~\cite{STAR:2021iop,STAR:2025zdq} at $\sNN \lesssim 20$~GeV relative to non-critical baselines. 

Interpreting experimental fluctuation measurements is a complex task due to several theoretical and methodological challenges. Grand canonical ensemble (GCE) calculations, commonly used to set the theoretical expectations, must be reconciled with the constraints imposed by exact baryon number conservation, finite-size effects, and nonequilibrium dynamics in heavy-ion collisions~\cite{Koch:2008ia, Berdnikov:1999ph}. Furthermore, experimental data are collected in momentum space, whereas theoretical predictions often focus on fluctuations in coordinate space. The extent to which critical fluctuations survive after transitioning from coordinate space to experimentally accessible momentum space remains an open question~\cite{Savchuk:2022msa, Vovchenko:2021kxx}.

Previous studies have employed molecular dynamics (MD) simulations of the Lennard-Jones (LJ) fluid as a microscopic model to investigate particle number fluctuations near the CP in a box setup~\cite{Kuznietsov:2022pcn}. These simulations, which were limited to second order fluctuations revealed significant fluctuations in coordinate space but showed that such fluctuations are largely washed out when analyzed in momentum space. This suppression arises because, in a uniform system with periodic boundary conditions, particle coordinates and momenta remain uncorrelated, causing the CP signal to vanish in momentum space.
However, coordinates and momenta are correlated in the presence of collective motion, which allows the CP signal to manifest in momentum space if the system is expanding~\cite{Kuznietsov:2023iyu}.

In this work, we extend our previous studies \cite{Kuznietsov:2022pcn, Kuznietsov:2023iyu} to non-Gaussian fluctuations, namely third and fourth-order cumulants in coordinate and momentum space acceptances. 
We study the equilibration dynamics of the different cumulants and identify the expected signals of the CP.
In addition to ordinary cumulants, we also study factorial cumulants, including the acceptance dependence of the recently suggested scaled factorial cumulants~\cite{Bzdak:2025rhp}. 
We also apply efficiency cuts to distinguish experimentally accessible protons from baryons, finding significant quantitative suppression of the CP signal in the former relative to the latter.

The paper is organized as follows: In Sec.~\ref{sec-setup}, we describe the molecular dynamics framework and introduce the observables we discuss and their main properties. Section~\ref{sec-thermalisation} studies system equilibration.
Sections~\ref{sec:Results_CS} and~\ref{sec:Results_MS} present our results for coordinate- and momentum-space fluctuations, including the effects of ensemble averaging, and discuss implications for heavy-ion collisions. In Sec.~\ref{sec:Summary}, we summarize our conclusions and outline directions for future research. Finally, in Appendix~\ref{KolafaEOS} we describe a derivation of the parametrized LJ EoS used in this article, and in Appendix~\ref{momentumIdG} we provide a method to derive ideal-gas cumulants in the microcanonical ensemble.

\section{Simulation setup}
\label{sec-setup}

\subsection{Lennard-Jones fluid}

The LJ fluid corresponds to a system of classical non-relativistic particles interacting via the following potential
\eq{
V_{\rm LJ}(r) = 4\varepsilon\left[\left(\frac{\sigma}{r}\right)^{12} - \left(\frac{\sigma}{r}\right)^{6}\right].
}
Here the first term corresponds to the repulsion at short distances while the second term models intermediate range attraction.
The two parameters -- $\sigma$ and $\varepsilon$ -- correspond to the size of the repulsive core and the depth of the attractive well, respectively, and define the corresponding length and energy scales in the system.

It is customary to use dimensionless variables by defining the reduced temperature $T^* = T / (k_B \varepsilon)$ and density $n^* = n \sigma^3$.
The particle mass defines the dimensionless time variable, $t^* = t \sqrt{\varepsilon/(m\sigma^2)}$.
Most properties of the LJ fluid, including the phase diagram in temperature-density plane, become independent of $\sigma$ and $\varepsilon$ in these variables.

Although the equation of state of LJ is not known exactly, it has been studied extensively with molecular dynamics simulations.
The phase diagram of the LJ fluid contains a rich phase structure, including a first-order liquid-gas phase transition with a CP in 3D-Ising universality class~\cite{STEPHAN2020112772}, located at $T_c^* = 1.321 \pm 0.007$ and $n_c^* = 0.316 \pm 0.005$~\cite{doi:10.1021/acs.jcim.9b00620}.
Furthermore, many parametrizations exist in the literature, which accurately describe molecular dynamics data on the LJ EoS in both gas and liquid phases~(see~Ref.~\cite{STEPHAN2020112772} for review).

\begin{figure*}[t]
    \includegraphics[width=\textwidth]{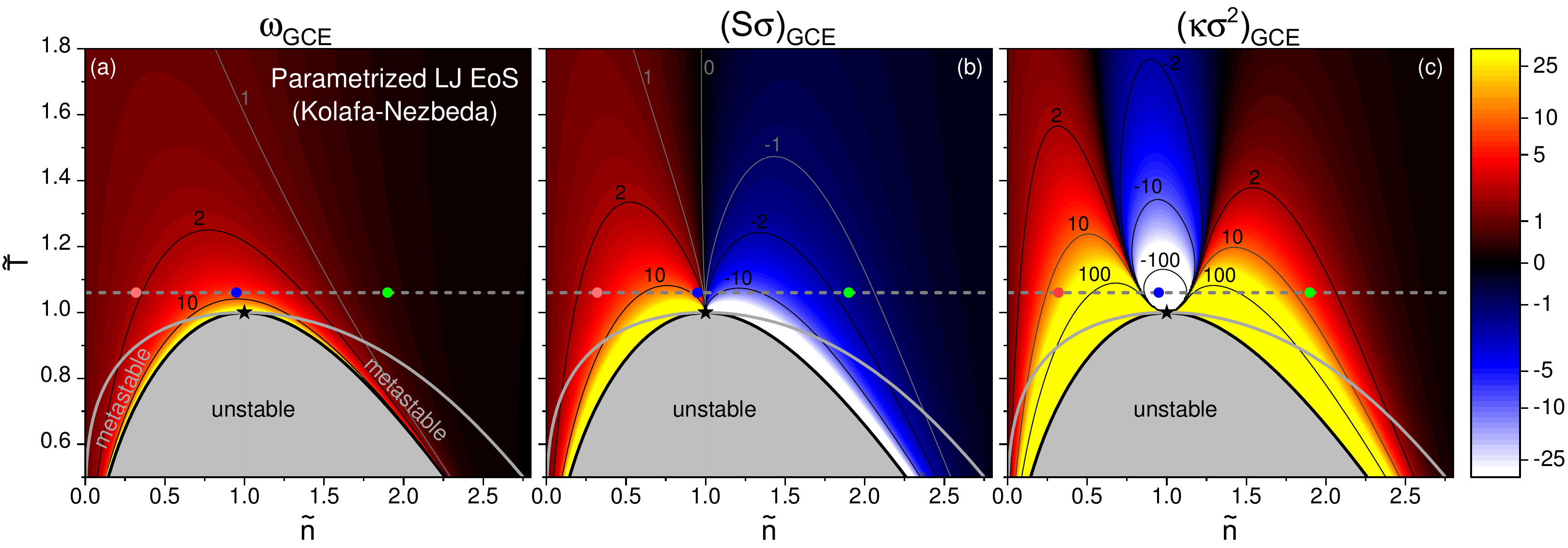}
    \caption{Scaled variance (a), skewness (b), and kurtosis (c) of grand-canonical particle number fluctuations in the Kolafa-Nezbeda equation of state (see Appendix \ref{KolafaEOS} for details) for the Lennard–Jones fluid in the $(\tilde{T},\tilde{n})$ plane.
The critical point is indicated by a black star, while the binodal and spinodal lines are shown by solid gray and solid black curves, respectively.
Contours of constant values for fluctuation measures are shown as thin lines.
The isotherm at $T = 1.06\,T_{\rm c}$ is represented by a dashed line, and the three densities along this isotherm selected for simulations, $n = 0.32\,n_{\rm c}$, $n = 0.95\,n_{\rm c}$, and $n = 1.9\,n_{\rm c}$, are marked by red, blue, and green circles, respectively. 
}
    \label{fig-fluks-gce}
\end{figure*}

\begin{figure*}[t]
    \includegraphics[width=0.94\textwidth]{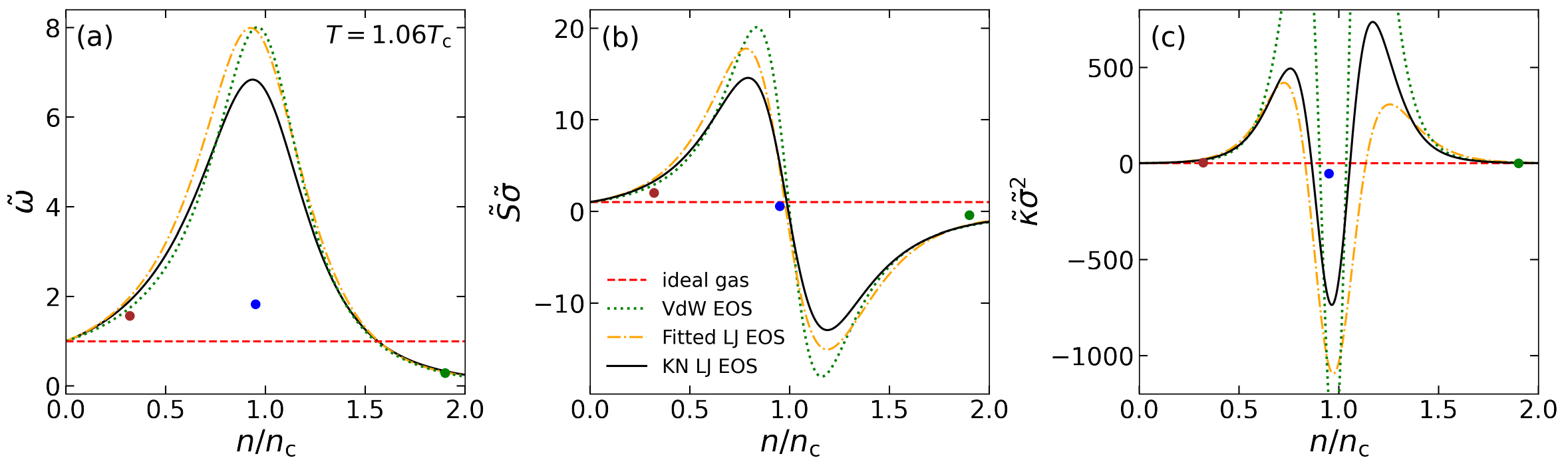}
    \caption{
    Scaled variance, skewness, and kurtosis of grand-canonical particle-number fluctuations in the Kolafa–Nezbeda (black solid curves) and virial equation of state (yellow dash-dotted curves). Van der Waals EoS results (green dotted curves) are provided for comparison.
    The dots represent the maximum value for $N=400$ LJ data with GCE correction (see Fig. \ref{fig-moments-alphadep}).
    \label{fig-virial-cumulants}}
\end{figure*}

\subsection{Grand-canonical expectations}

Before proceeding to molecular dynamics simulations, we first discuss the GCE expectations in the thermodynamic limit (TDL).
Fluctuations of particle number are our main observable of interest.
In the GCE, these fluctuations correspond to susceptibilities $\chi_n = [\partial^n (p/T) / \partial (\mu/T)^n ]_{T, V}$ as
\eq{\label{eq:pres}
\kappa_n = V T^3\, \chi_n,
}
where $\kappa_n$ is the $n$th order cumulant and $V$ is the system volume and $\mu$ corresponds to conserved number of baryons.

One can construct ratios of cumulant to obtain intensive (volume-independent) fluctuation measures such as the scaled variance $\omega_{\rm GCE}$, skewness $(S\sigma)_{\rm GCE}$, and excess kurtosis $(\kappa\sigma^2)_{\rm GCE}$
\eq{
\label{eq:omega}
\omega_{\rm GCE} = \frac{\chi_2}{\chi_1},\\
\label{eq:ssigma}
(S\sigma)_{\rm GCE} = \frac{\chi_3}{\chi_2},\\
\label{eq:ksigma}
(\kappa\sigma^2)_{\rm GCE} = \frac{\chi_4}{\chi_2}.
}
where the TDL is assumed with $N\to\infty, ~V\to \infty$.

To calculate the behavior of these quantities on the phase diagram of the LJ fluid we use a Kolafa–Nezbeda (KN) parametrization~\cite{KOLAFA19941} for the LJ equation of state (see Appendix~\ref{KolafaEOS}).
As discussed in Ref.~\cite{STEPHAN2020112772}, this parametrization yields highly accurate description of MD data as well as a reasonable description of the spinodal region.
One should note that this parametrization (as well as the vast majority of other parametrizations in the literature~\cite{STEPHAN2020112772}) is fully analytic and exhibits classical (mean-field) scaling behavior as opposed to 3D-Ising scaling. 
The critical point in the KN EoS is located at $T_c^* \simeq 1.340$ and $n_c^* \simeq 0.311$, with the temperature being somewhat higher than the consensus estimate ($T_c^* = 1.321 \pm 0.007$) for MD simulations.

Figure~\ref{fig-fluks-gce} presents the phase diagram in reduced variables $\tilde T=T^*/T_{\rm c}^*=T/T_{\rm c}$ and $\tilde n=n^*/n_{\rm c}^*=n/n_{\rm c}$.  
It shows $\omega_{\rm GCE}$, $(S\sigma)_{\rm GCE}$, and $(\kappa\sigma^2)_{\rm GCE}$ for grand-canonical particle-number fluctuations in the $(\tilde T,\tilde n)$ plane for the KN EoS. All three observables diverge at the critical point (CP) and remain enhanced well away from it. As with the scaled variance, the skewness and kurtosis diverge at the CP, but they exhibit stronger, path-dependent structure: $(S\sigma)_{\rm GCE}$ and $(\kappa\sigma^{2})_{\rm GCE}$ can be large and of either sign, depending on the approach to the CP.
Qualitatively, the structure is identical to that in the van der Waals EoS~\cite{Vovchenko:2015xja}.

The isotherm at $T = 1.06\,T_{\rm c}$, considered in our MD simulations, is shown by the dashed horizontal line in Fig. \ref{fig-fluks-gce}.
We simulate three densities, $n = 0.32\,n_{\rm c}$, $0.95\,n_{\rm c}$, and $1.9\,n_{\rm c}$, which are marked by red, blue, and green circles, respectively. 
One sees that all three densities correspond to strongly non-Gaussian fluctuations: $\omega$ is large in the dilute phase and peaks near the CP, while it drops below unity in the dense phase; $S\sigma$ is large and positive in the dilute phase, large and negative in the dense phase, and small near the critical density at $\tilde T>1$; $\kappa\sigma^2$ is strongly positive in both dilute and dense phases but becomes strongly negative in the immediate vicinity of the CP.

The dependence of $\tilde \omega$, $\tilde S\tilde \sigma$, and $\tilde \kappa \tilde \sigma^2$ on $\tilde n$ along the $\tilde T = 1.06$ isotherm is shown in Fig.~\ref{fig-virial-cumulants} for the Kolafa–Nezbeda EoS, the virial LJ EoS (cf. Ref.~\cite{Kuznietsov:2022pcn}) and the classical van der Waals EoS~\cite{Vovchenko:2015uda} by solid black, dash-dotted yellow, and dotted green lines, respectively. The GCE Boltzmann ideal-gas baseline, $\tilde \omega = \tilde S\tilde \sigma = \tilde \kappa\tilde \sigma^2 = 1$, is indicated by red dashed horizontal lines. Significant deviations from the baseline are observed for all three equations of state: $\omega$ exhibits a peak near $n_c$, $S\sigma$ shows both a peak and a trough, and $\kappa\sigma^2$ displays a double-peak structure with an intervening trough in the vicinity of $n_c$. Such behavior of higher-order cumulants is a well-recognized hallmark of the critical point.
Although the cumulants exhibit qualitatively similar behavior across the three parameterized EoS, their values differ quantitatively, and the mismatch increases with cumulant order.

\subsection{Molecular dynamics}

\subsubsection{Simulation setup}

MD simulations proceed by numerically integrating Newton's equations of motion.
The simulations are performed using the Velocity-Verlet integration method for a system of $N$ particles with periodic boundary conditions in the minimum-image convention\footnote{Details of the method can be found in \cite{Allen2017}; the simulation setup source code is available in \cite{LJgithub}}.
In our first work~\cite{Kuznietsov:2022pcn}, we used the simulations to study the behavior of particle-number fluctuations along the supercritical isotherm $\tilde T = 1.06 \tilde T_c$.
This was achieved by performing the simulations for a long period of time at each value of the particle number density and computing the moments of the particle number distribution as time averages.
This analysis was extended to ensemble averaging in Ref.~\cite{Kuznietsov:2024xyn}, where we also incorporated longitudinal collective flow, introducing correlations between particle coordinates and momenta.

In the present work, we explore the same conditions of temperature and density as in Refs.~\cite{Kuznietsov:2022pcn, Kuznietsov:2024xyn} and use the same GPU-accelerated MD simulation code from~\cite{LJgithub}.
We refer to Sec. III of Ref.~\cite{Kuznietsov:2022pcn} for the details of MD simulation framework.
The key new element of the present work is that we study higher-order cumulants, including third- and fourth-order moments in coordinate and momentum space acceptances, by significantly increasing the number of simulated events at each density.

Our simulations here are performed for $N = 400$, which approximately corresponds to the total number of baryons in central collisions of heavy ions when the production of baryon-antibaryon pairs is negligible.

\subsubsection{External conditions}

We perform simulations at three points in the phase diagram. 
They all correspond to the same temperature of $\tilde T = 1.4 \simeq 1.06 \, \tilde T_c$ but different values of the number density: (i) $\tilde n = 0.1 \simeq 0.32 \, \tilde n_c$~(dilute), (ii) $\tilde n = 0.3 \simeq 0.95 \, \tilde n_c$~(near critical), and (iii) $\tilde n = 0.6 \simeq 1.90 \, \tilde n_c$~(dense).
These three cases probe different regimes with respect to the interaction, namely one dominated by attraction (dilute), critical effects (near critical), and repulsion (dense).
The value of the density determines the length of the simulation box, $\tilde L = (N / \tilde n)^{1/3}$, where $N = 400$.
The simulations are performed in the microcanonical ensemble, where the energy per particle $\tilde u = \tilde U / N$, rather than the temperature $\tilde T$, is a fixed quantity strictly conserved throughout the evolution.
To achieve the desired mapping of the microcanonical simulation to the proper $(\tilde T, \tilde n)$ point on the phase diagram, we initialize the system with the energy per particle $\tilde u$ that matches the value from the LJ equation of state.
We cross-check that the average value of the kinetic temperature during the simulation matches $T = 1.06T_{\rm c}$ to a relative accuracy of about 1$\%$ once equilibrium is reached.

\subsection{Observables}
\label{sec:analysis}

For each $i$th simulated event, the number of particles within the desired acceptance is computed and added to the statistics.
The values of the (factorial) cumulants and their statistical errors are computed using the \texttt{sample-moments} package~\cite{sample-moments}, which implements the Delta method for the estimation of the standard errors.

\subsubsection{Ordinary cumulants}

The key fluctuation observables studied in this work are the scaled variance $\omega$, skewness $S\sigma$, and excess kurtosis $\kappa\sigma^2$, defined in Eqs.~\eqref{eq:omega}, \eqref{eq:ssigma}, and \eqref{eq:ksigma}, respectively. In a finite-size system, one would rather use the statistical definition of central moments $\kappa_n$,
\eq{\begin{split}
    \omega = &\frac{\kappa_2}{\kappa_1} =\frac{\mean{(\Delta N)^2}}{\mean{N}}, ~~S\sigma = \frac{\kappa_3}{\kappa_2} = \frac{\mean{(\Delta N)^3}}{\mean{(\Delta N)^2}},\\ &\kappa\sigma^2= \frac{\kappa_4}{\kappa_2} =\frac{\mean{(\Delta N)^4}}{\mean{(\Delta N)^2}} - \mean{(\Delta N)^2}.
\end{split}}
where $\mean{\dots}$ corresponds to ensemble averaging in a coordinate- or momentum-space acceptance in a system with conserved baryon number.

A Subensemble Acceptance Method (SAM) was developed to account for this effect in Ref.~\cite{Vovchenko:2020tsr}, which provides the relation between the cumulants in the (micro)canonical and grand-canonical ensemble. We use the SAM to reconstruct the expected cumulant ratios in the grand-canonical ensemble from the cumulants in the microcanonical ensemble obtained from the simulations:
\begin{align}
    \tilde{\omega} &= \frac{\omega}{1 - \alpha}, \label{eq:wtil}\\
    \tilde{S} \tilde{\sigma} &= \frac{S\sigma}{1 - 2\alpha},  \label{eq:ssigmgce}\\
    \tilde{\kappa} \tilde{\sigma}^2 &= \frac{\kappa\sigma^2}{1 - 3\alpha (1 - \alpha)} + \frac{3\alpha (1 - \alpha) (\tilde{S}\tilde{\sigma})^2}{1 - 3\alpha (1 - \alpha)},\label{eq:kurtgce}
\end{align}
where $\alpha = \langle N_{\rm acc} \rangle / N$ is the acceptance fraction. 
These relations are expected to be exact only in the thermodynamic limit, $N \to \infty$, and only for acceptance cuts in coordinate space. Note that due to division by zero and finite precision at $\alpha = 1/2$ for skewness and kurtosis, the practical applicability of the corrections becomes limited close to $\alpha = 1/2$. This is discussed in detail in Sec. \ref{sec:Results_CS}, B.

\begin{figure*}[t]
    \includegraphics[width=0.95\textwidth]{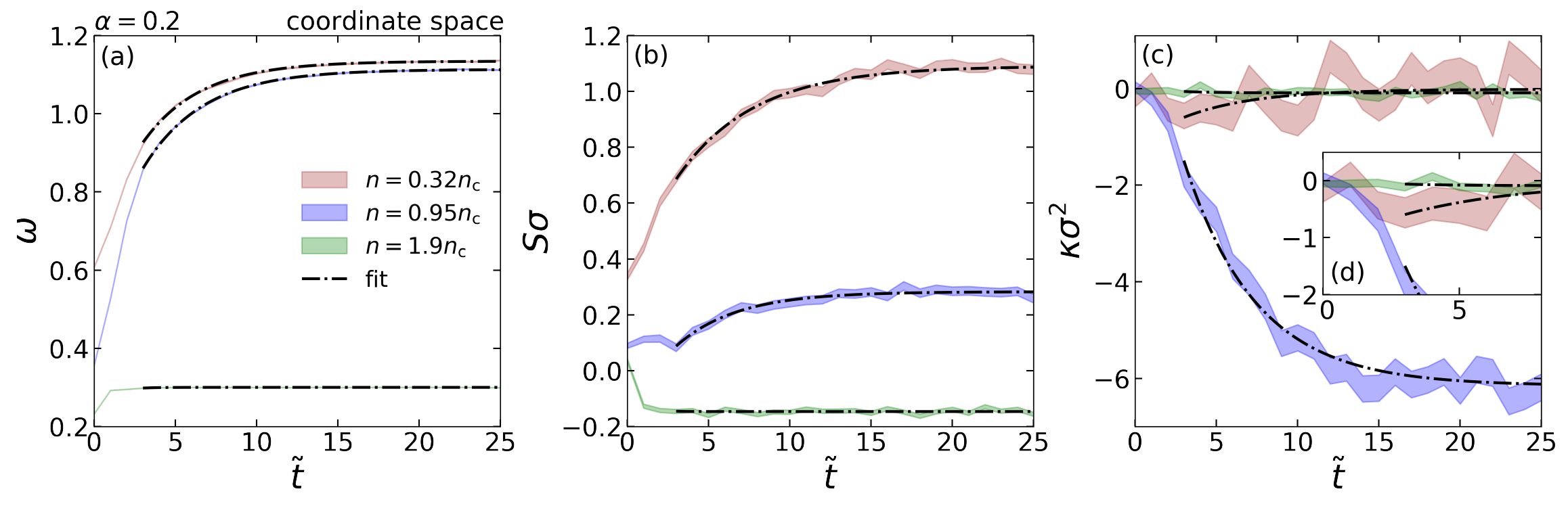}
    \caption{
    Scaled variance (a), skewness (b), and kurtosis (c) of particle-number fluctuations in the Lennard–Jones fluid within a coordinate-space subsystem with $\alpha=0.2$ at temperature $T = 1.06T_{\rm c}$ as a function of the reduced time $\tilde t$. Three values of density, $n=0.32 n_{\rm c}$, $n=0.95 n_{\rm c}$, and $n=1.9 n_{\rm c}$, are presented by, respectively, red, blue, and green bands. The widths of the bands represent statistical uncertainties obtained using the Delta method. The inset (d) highlights the early-time behavior of the kurtosis. The dot-dashed lines correspond to the parametric fit by exponential relaxation~(\ref{eq:fit}) with parameter values from Tab.~\ref{tab:exp}.
    }
    \label{fig-timedep}
\end{figure*}

\subsubsection{Factorial cumulants}

Factorial cumulants $\hat C_n$ provide complementary information to the standard cumulants $\kappa_n$ and they are particular useful for probing multi-particle correlations~\cite{Bzdak:2016sxg}.
They can be expressed as linear combinations of the standard cumulants, that subtract all low-order correlations:
\begin{align}\label{eq:fc2}
    \hat C_2 &=- \mean{N} + \kappa_2, \\\label{eq:fc3}
    \hat C_3 &= 2\mean{N} - 3 \kappa_2 + \kappa_3, \\\label{eq:fc4}
    \hat C_4 & =  -6\mean{N} + 11\kappa_2 - 6\kappa_3 + \kappa_4
\end{align}

The factorial cumulants are normalized by the mean number of particles $\mean{N}$ to yield the normalized factorial cumulants $\hat C_n / \mean{N}$, and these quantities have been presented by STAR~\cite{STAR:2020tga,STAR:2025zdq}.

In addition, we also explore scaled factorial cumulants defined as $\hat c_n = \hat C_n / \mean{N}^n$~\cite{Bzdak:2016sxg}.
As discussed in Ref.~\cite{Bzdak:2025rhp}, the acceptance dependence of the scaled factorial cumulants is expected to distinguish between short- and long-range correlations.

\section{Equilibration}
\label{sec-thermalisation}

\begin{figure*}[t]
    \includegraphics[width=0.95\textwidth]{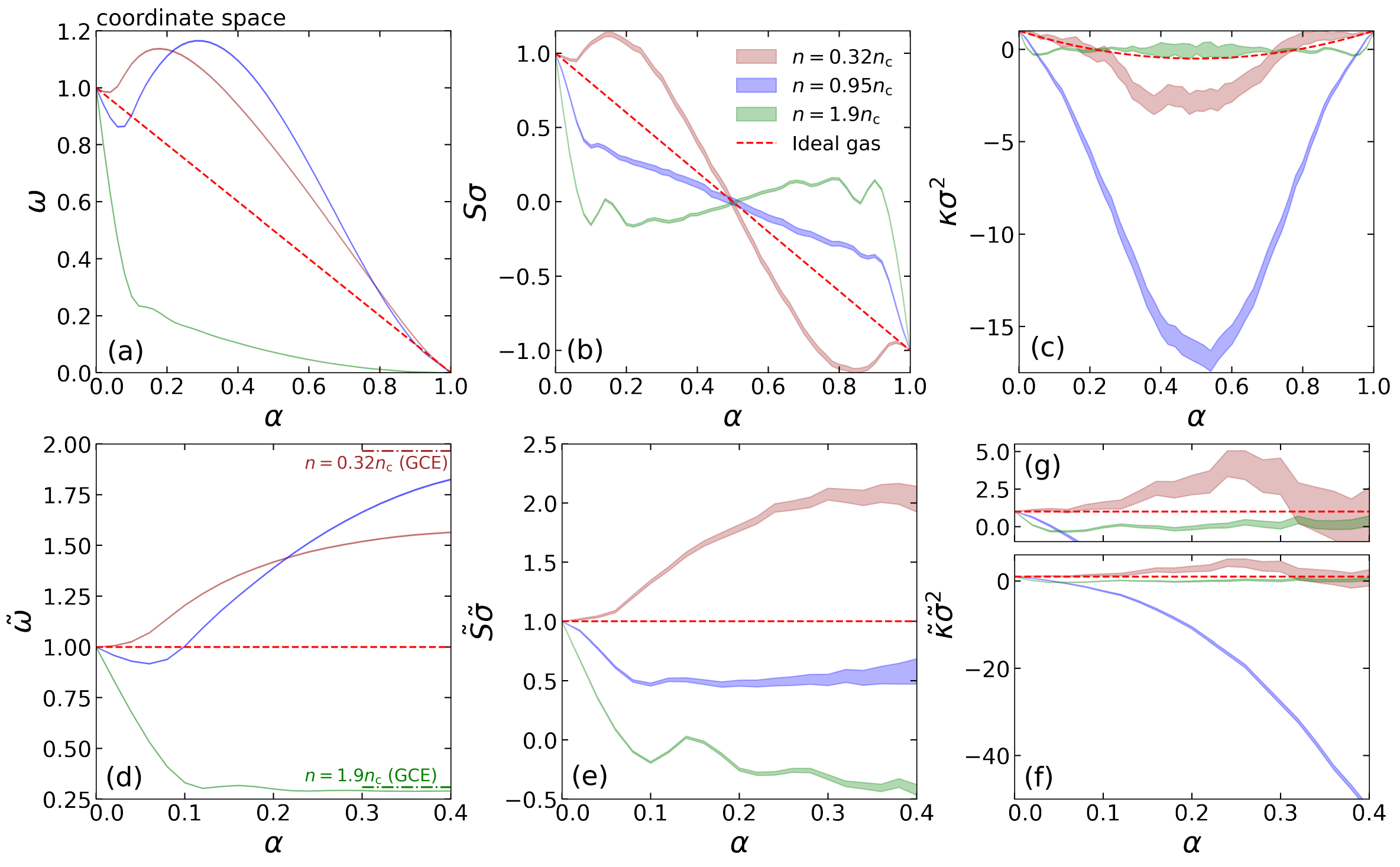}
    \caption{
    The same as in Fig.~\ref{fig-timedep} but for scaled variance (a), skewness (b), and kurtosis (c) as functions of acceptance fraction $\alpha$ after thermalization. Panels (d), (e), and (g) show the corresponding fluctuation measures corrected for global baryon number conservation using Eqs.~(\ref{eq:wtil})-(\ref{eq:kurtgce}), respectively.
    The dash-dotted lines in (d) show the expected result in the thermodynamic limit.
    Additionally, panel (g) shows a zoomed-in view of the kurtosis behavior for densities far from the critical point.
    The dashed red lines represent the ideal gas baseline.
    }
    \label{fig-moments-alphadep}
\end{figure*}

We begin our analysis by studying the time dependence of fluctuations in coordinate space.
Figure~\ref{fig-timedep} presents scaled variance~(\ref{eq:omega}), skewness~(\ref{eq:ssigma}), and kurtosis~(\ref{eq:ksigma}) as functions of reduced time $\tilde t$ computed for acceptance fraction $\alpha=0.2$.
The temperature is fixed to $T=1.06T_c$ and the three values of particle number density described earlier are considered.
The procedure used to calculate time dependent cumulants is described in Ref.~\cite{Kuznietsov:2024xyn}. 
The obtained time evolution reveals a clear relaxation behavior toward equilibrium for all three fluctuation measures.
The results are qualitatively similar for all acceptance fractions $\alpha$.

To extract the equilibration time for different cumulants at different densities, we fit the time-dependence from Fig.~\ref{fig-timedep} 
using the standard exponential relaxation formula:
\eq{
K(\tilde t) = K_{\infty} + A e^{-\tilde t/\tau}, \label{eq:fit}
}
where $K$ stands for the fluctuation measure under consideration, $\omega$, $S\sigma$, or $\kappa \sigma^2$.
The asymptotic value $K_{\infty}$ represents the equilibrium value of the fluctuation measure,
$\tau$ is the relaxation time, and $A$ is a phenomenological constant.
We extract the values and uncertainties of these parameters through a $\chi^2$ fit.

\begin{table}
\label{tab:exp}
\begin{tabular}{c|c|c|c}
 & $\omega$ & $S \sigma$ & $\kappa \sigma^2$ \\
\hline
\hline
$n/n_{\rm c}$ & \multicolumn{3}{|c}{$0.32$} \\
\hline
$K_{\infty}$  & $ 1.13 \pm 0.01$  & $1.09 \pm 0.01$  & $-0.02 \pm 0.06$ \\
\hline
$\tau$        & $3.56 \pm 0.03$  & $4.78 \pm 0.32$  & $4.38 \pm 3.84$  \\
\hline
\hline
$n/n_{\rm c}$ & \multicolumn{3}{|c}{$0.6$} \\
\hline
$K_{\infty}$  & $ 1.30 \pm 0.01$  & $1.04 \pm 0.01$  & $ -3.74 \pm 0.1$ \\
\hline
$\tau$        & $3.45 \pm 0.02$  & $4.01 \pm 0.36$  & $4.25 \pm 1.7$  \\
\hline
\hline
$n/n_{\rm c}$ & \multicolumn{3}{|c}{$0.95$} \\
\hline
$K_{\infty}$  & $1.12 \pm 0.01$  & $0.28 \pm 0.01$  & $-6.15 \pm 0.09$ \\
\hline
$\tau$        & $3.70 \pm 0.02$  & $3.75 \pm 0.47$  & $4.45 \pm 1.41$  \\
\hline
\hline
$n/n_{\rm c}$ & \multicolumn{3}{|c}{$1.9$} \\
\hline
$K_{\infty}$  & $0.30 \pm 0.01$  & $-0.15 \pm 0.01$ & $-0.09 \pm 0.02$ \\
\hline
$\tau$        & $0.98 \pm 0.14$  & $0.63 \pm 3.20$  & $1.97 \pm 4.78$  \\
\hline
\end{tabular}
\caption{Asymptotic values $K_{\infty}$ and relaxation timescales $\tau$ for different cumulant ratios, obtained by fitting the exponential relaxation function~\eqref{eq:fit} to the time evolution of fluctuation measures shown in Fig.~\ref{fig-timedep}.}
\end{table}

The fitted values of parameters $K_{\infty}$ and $\tau$ are presented in Tab.~\ref{tab:exp}. 
For the scaled variance $\omega$ we observe the largest relaxation time among the three densities in the vicinity of the critical point ($n=0.95 n_{\rm c}$).
The relaxation time is comparably large in the dilute phase ($n=0.32 n_{\rm c}$), where particle collisions are relatively rare and the mean free path is long.
Correspondingly, the relaxation time is smallest in the dense phase ($n=1.9 n_{\rm c}$) where the mean free path is short.
The fact that the relaxation time is non-monotonic with respect to density, with a local peak near the critical density ($n=0.95 n_{\rm c}$),
represents the effect of critical slowing down.

For higher-order cumulants, the statistical errors are larger and the extracted relaxation times have larger uncertainties. The presence of a local peak near the critical density ($n=0.95 n_{\rm c}$) is less clear.
However, one can observe that the relaxation times for $S\sigma$ and $\kappa\sigma^2$ are of comparable magnitude to $\omega$ near the critical point ($n=0.95 n_{\rm c}$).
This indicates that the cumulants of different order near the CP may equilibrate at comparable timescales
which is consistent with results obtained within stochastic fluid dynamics~\cite{Chattopadhyay:2024jlh}.

The obtained time dependencies confirm that the system reliably reaches equilibrium values $K_{\infty}$ even with respect to kurtosis and in the vicinity of the CP at a sufficiently large time $\tilde t\gtrsim 25$.
In the following we study the system at $\tilde t = 50$ assuming its full thermalization.

\section{Fluctuations in Coordinate Subspace}
\label{sec:Results_CS}

\subsection{Coordinate Space Cumulants}
Let us now examine the fluctuations in coordinate space as a function of the acceptance fraction $\alpha$ after thermalization.
Panels (a)-(c) of Fig. \ref{fig-moments-alphadep} present $\alpha$-dependence of scaled variance~[Eq.~\eqref{eq:omega}], skewness~[Eq.~\eqref{eq:ssigma}], and kurtosis~[Eq.~\eqref{eq:ksigma}], respectively, before baryon number conservation corrections.
The red dashed lines show the non-interacting (ideal-gas) baseline, given by the binomial distribution.
When $\alpha \to 0$, all three fluctuation measures approach unity, reflecting the expected Poisson limit in small acceptance.
In the opposite limit, $\alpha \to 1$, exact particle-number conservation suppresses fluctuations and the cumulant ratios approach their full-acceptance limits (namely $\omega \to 0,~ S\sigma \to -1,~ \kappa \sigma^2 \to 1$). Both limits are reproduced in the simulations in Fig. \ref{fig-moments-alphadep} (a,b,c).

The scaled variance [Fig. \ref{fig-moments-alphadep}, panel (a)] shows enhancement of fluctuations near the CP in the intermediate-$\alpha$ regime; however, this enhancement is significantly suppressed by global conservation and finite size effects. 
A strong suppression of fluctuations is observed in the dense phase ($n = 1.9 n_{\rm c}$), reflecting the dominance of repulsive interactions.
These results reproduce our earlier simulations, and a detailed discussion of second-order cumulants can be found in Refs.~\cite{Kuznietsov:2022pcn, Kuznietsov:2024xyn}.

The skewness [Fig. \ref{fig-moments-alphadep}, panel (b)] is antisymmetric around $\alpha = 0.5$ for all densities, namely $S\sigma(\alpha) = -S\sigma(1-\alpha)$ and $S\sigma = 0$ at $\alpha = 0.5$. This reflects the symmetry between the exchange of the subsystem and its complement~\cite{Bzdak:2017lt}.
Focusing on $\alpha < 0.5$, one observes that in the dilute system $S\sigma$ is above the ideal-gas baseline, whereas near the CP and in dense systems the skewness is suppressed relative to the ideal gas.
This qualitative evolution with the density across a supercritical isotherm reflects the transition from dilute to dense phase near a CP.

The kurtosis [Fig. \ref{fig-moments-alphadep}, panel (c)] shows strong sensitivity to the CP proximity.
For instance, at $n = 0.95n_{c}$ the kurtosis attains large negative values, with $\kappa\sigma^{2} \simeq -17.5$ reached for $\alpha = 0.5$.
On the other hand, the kurtoses in the dilute and dense phases are relatively closer to the ideal gas baseline.  
The dilute case ($n=0.32 n_{\rm c}$) shows similar $\alpha$-dependence but with significantly smaller magnitude of the kurtosis.
For the dense case ($n=1.9 n_{\rm c}$), the kurtosis values are close to zero and no notable $\alpha$ dependence is observed within statistical errors.

Large absolute values of kurtosis and change in the sign are typical behaviors in the vicinity of the critical point. This indicates that higher-order cumulants are useful in probing the phase diagram for the CP signals even in the presence of strong canonical ensemble effects.

One can see that the presented cumulant ratios show non-monotonic behavior at small $\alpha \lesssim 0.2$.
We attribute this behavior to finite particle number statistics and the fact that the longitudinal size of the subsystem for small $\alpha$ is comparable to the size of a single LJ particle.
These oscillations are the most apparent in the dense phase ($n = 1.9 n_{\rm c}$), where the physical size of the simulation box is the smallest, for the skewness.
This phenomenon has been discussed in more detail in Ref.~\cite{Poberezhnyuk:2020ayn}.

\subsection{Correction for baryon conservation}

Equations~\eqref{eq:wtil}, \eqref{eq:ssigmgce}, and \eqref{eq:kurtgce} express the cumulant ratio inside the coordinate subsystem in terms of GCE ratios.
These expressions are obtained in the thermodynamic limit, $N \to \infty$.
Here we use these relations to extract the GCE cumulant ratios from our MD simulations. 
Of course, the actual GCE values are only expected to be recovered in the thermodynamic limit, while our simulations are done for $N = 400$.
Therefore, distortions are expected, especially in vicinity of the CP where finit-size effects are expected to be strong.

The corrected fluctuation measures $\tilde{\omega}$, $\tilde{S}\tilde \sigma$, and $\tilde{\kappa}\tilde \sigma^2$ are presented in Fig.~\ref{fig-moments-alphadep} panels (d), (e), and (f), respectively.
We show the results in a restricted acceptance range of \( 0 \leq \alpha \leq 0.4 \).
The reason is that the analytical correction involves a pole at $\alpha = 0.5$. This pole does not have physical significance, but it does lead to a large increase in statistical error.
Also, the results at $\alpha > 0.5$ are connected to those at $\alpha < 0.5$ by symmetry and thus contain no additional information.
For these reasons, we restrict our analysis of the corrected cumulant ratios to $\alpha \in [0,0.4]$. 

Panel (d) of Fig.~\ref{fig-moments-alphadep} depicts scaled variance, corrected for baryon number conservation. 
One can see that for an almost critical density of \( n=0.95n_{\rm c} \) scaled variance demonstrates a significantly stronger acceptance behavior compared to \( n=0.32n_{\rm c} \), indicating that 
the assumption of the thermodynamic limit
breaks down for the given system size. 
On the other hand, for \( n=0.32n_{\rm c} \) and \( n=1.9n_{\rm c} \), where the correlation length remains smaller,
\( \tilde{\omega} \) reaches plateau at $\alpha \gtrsim 0.4$.
The corresponding values $\tilde{\omega}\approx 1.52$ for \( n=0.32n_{\rm c} \) and $\tilde{\omega}\approx 0.25$ for \( n=1.9n_{\rm c} \) in $\alpha=0.4$ can be compared with the GCE estimates of the parametrized Lennard-Jones EOS. 
The latter yields $\omega_{\rm gce}= 1.9$~($n = 0.32n_{\rm c}$), $6.75$~($n = 0.95n_{\rm c}$) and $0.33$~($n = 1.9n_{\rm c}$), comparable to the values extracted from MD simulations for dense system only. For dilute system and near to the CP density, $\tilde{\omega}$ can be interpreted only as a lower bound on the true GCE value in thermodynamic limit.

The corrected skewness is depicted in panel (e) of Fig.~\ref{fig-moments-alphadep}.
Despite larger uncertainties, its values saturate for \( n=0.32n_{\rm c} \) and \( n=0.95n_{\rm c} \) at \( \alpha \gtrsim 0.25 \) and \( \alpha \gtrsim 0.1 \), respectively.
The corresponding values are positive, $\tilde{S}\tilde{\sigma} \approx 2$, in the dilute region and negative, $\tilde{S}\tilde{\sigma} \approx -0.5$ in the dense region. 
This is consistent with the estimates from the parametrized LJ EoS and with qualitative expectations.
From a quantitative point of view, however, the simulation results significantly underestimate the GCE predictions [$(S\sigma)_{\rm gce}=3.27$ and $(S\sigma)_{\rm gce}=-1.53$ for $n=0.32n_{\rm c}$ and $n=0.95n_{\rm c}$, respectively] in magnitude. 
This indicates that finite-size effects in the third-order cumulant are strong for an LJ system of \( N = 400 \) particles and the thermodynamic limit for this quantity is still far away.
It is notable as well that $\tilde{S}\tilde{\sigma}$ approaches a plateau also for $n=0.95 n_c$, what is consistent with Fig. \ref{fig-virial-cumulants}, (b) curve near the CP.

Panel (f) of Fig. \ref{fig-moments-alphadep} presents the corrected kurtosis. Notably, for low and high densities, the kurtosis approaches a plateau (see Fig.~\ref{fig-moments-alphadep} panel (g) for more details). However, near the CP \( (0.95n_{\rm c}, 1.06T_{\rm c}) \), large negative values of kurtosis are observed, with a strong dependence on \( \alpha \). 
While these values are still far away from the GCE limit due to strong finite-size effects, our simulations support the idea that an observation of large negative kurtosis may potentially serve as a signal of the critical point nearby~\cite{Stephanov:2011pb}.

\begin{figure}
    \includegraphics[width=0.60\textwidth]{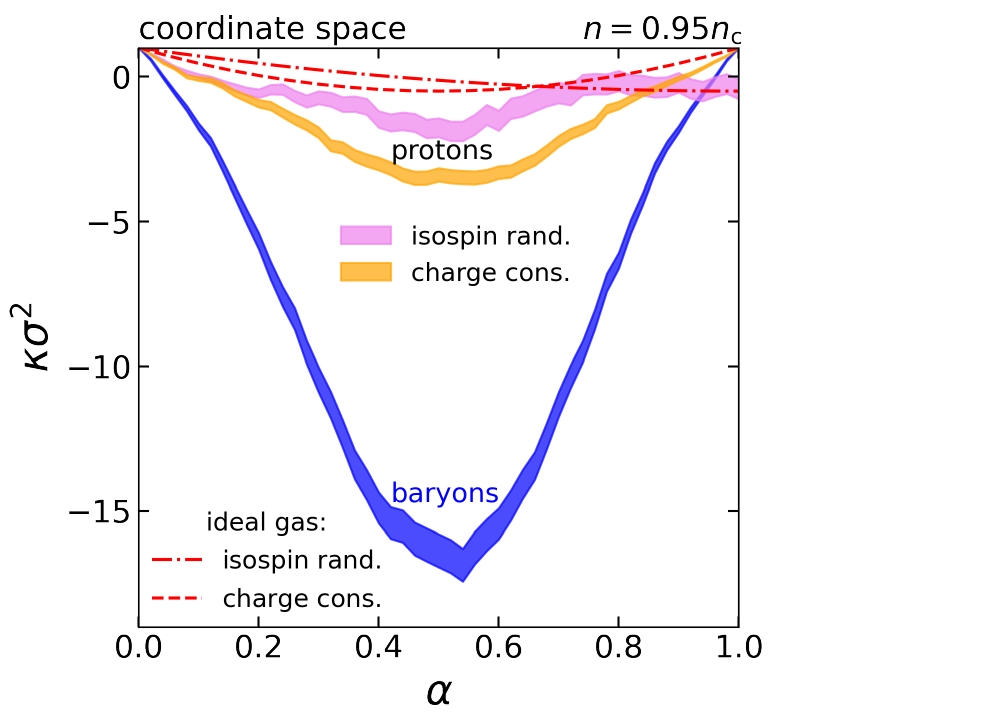}
    \caption{
    $\alpha$-dependence of kurtosis $\kappa\sigma^2$ in the vicinity of the critical point of Lennard–Jones fluid, $T = 1.06T_{\rm c}$ and $n = 0.95n_{\rm c}$ (middle) evaluated within a coordinate-space subsystem at $\tilde t = 50$. Baryon-number fluctuations are shown as blue bands, while proton-number fluctuations with isospin randomization and with additional charge conservation are represented by pink and yellow bands, respectively. The corresponding ideal-gas baselines are indicated by dashed (isospin randomization) and dash-dotted (charge conservation) lines. Note that ideal gas baseline is the same for baryons and protons under charge conservation scenario.
    }
    \label{fig-prot-standcumulants}
\end{figure}

\subsection{Proton vs baryon}
\label{sec-conservation}

\begin{figure*}
    \includegraphics[width=0.9\textwidth]{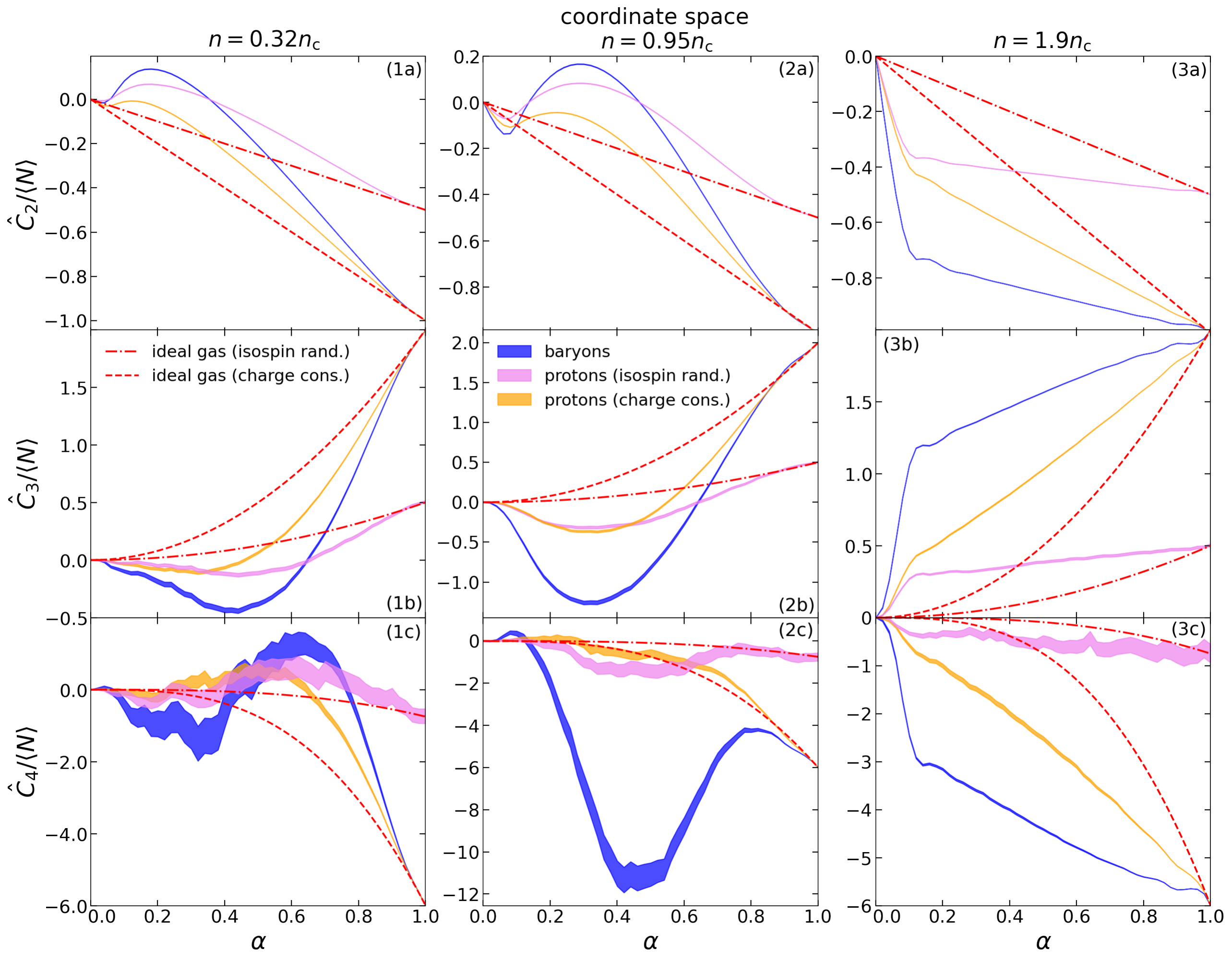}
    \caption{
   The same as in Fig. \ref{fig-prot-standcumulants} but for normalized factorial cumulants $\hat{C}_2/\langle N\rangle$ (top), $\hat{C}_3/\langle N\rangle$ (middle), and $\hat{C}_4/\langle N\rangle$ (bottom).
    }
    \label{fig-factmoments}
\end{figure*}

In the context of heavy-ion collisions, the particles in our simulation correspond to baryons.
However, experimental measurements are restricted to charged particles and cannot measure all baryons (such as neutrons).
For this reason, the proton number is commonly used as a proxy for the baryon number.
We thus also evaluate proton number fluctuations in our simulations by considering the protons as a subset of all particles in the simulation.
We consider two scenarios, expected to be relevant at high and low collision energies, respectively.
\begin{itemize}    
\item 
    \textbf{Isospin randomisation:}
    We label each baryon as a proton with a probability of 50\%, independently for each baryon.
    This corresponds to the complete randomisation of the isospin,
    expected to be accurate for large collision energies ($\sNN\gtrsim 10$ GeV), where hadronic rescattering involving pions equilibrates the isospin composition among baryons. 
    In this case, factorial cumulants are expected to scale as $\hat{C}_k^{\rm B} = 2^{k} \hat{C}_k^{\rm p}$ for an isospin-symmetric system~\cite{Kitazawa:2012at} where $\hat{C}_k^{\rm B}$ and $\hat{C}_k^{\rm p}$ are $k$-th order factorial cumulants for baryons and protons, respectively.

\item 
    \textbf{Charge conservation (conserved proton number):} 
    Pion production is suppressed at low collision energies ($\sNN < 10$ GeV), and the majority of the electric charge is carried by the protons in the final state.
    The global electric charge conservation thus implies that the total number of protons is fixed in this limit. 
    To implement this scenario, we identify the first $N/2 = 200$ particles in our simulation as protons at initialization and track the fluctuations of their number inside a subvolume throughout the simulation.
    
\end{itemize}

To illustrate the difference between baryon and proton cumulants, we show in Fig.~\ref{fig-prot-standcumulants} the $\alpha$-dependence of the (uncorrected) kurtosis $\kappa\sigma^{2}$ in the vicinity of the CP ($T=1.06T_c$ and $n=0.95n_c$) in a coordinate–space acceptance. 
The results for baryons are plotted in blue, while the ones for protons are shown for two sampling schemes: isospin randomization (violet) and global charge conservation (yellow). The corresponding ideal gas baselines are indicated by red dashed (isospin randomization: only baryon number conserved) and red dash-dotted (charge conservation: baryon and electric charge conserved) lines. As expected, all curves approach the Poisson limit as $\alpha\to 0$.
In the limit $\alpha \to 1$, the baryon kurtosis as well as proton kurtosis in the charge conservation scenario approach unity reflecting their exact conservation in full space.
In the isospin randomization case, however, the limit of unity is not reached for protons, because their number is not conserved even in full space.

One can see a pronounced CP signal in the kurtosis of baryon number, showing significantly negative values in a broad range of $\alpha$ values.
In contrast, the signal for the protons is significantly diluted in both scenarios.

Calculations in the isospin randomization~(violet) scenario proceed through random tagging of each baryon as a proton with probability $p\simeq 1/2$, corresponding to a binomial thinning of the baryon number distribution. 
This leads to a Poissonization of fluctuations, which is more prominent in high-order cumulants.
Note that even the ideal gas (IdG) results are affected, where Poissonization reduces the effect of baryon conservation.

Let us compare the deviations of the proton number kurtosis from the ideal gas baseline in the isospin randomization and charge conservation scenarios.
One can see that the deviation is larger in the charge conservation case.
This can be understood in terms of factorial cumulants of different orders.
As shown in Fig. \ref{fig-factmoments}, the deviations from the IdG are identical for  second-order 
factorial cumulants, but are 
stronger in the charge conservation case for the third-order factorial cumulants.
Given that the kurtosis contains contributions from all lower-order factorial cumulants,
\eq{\kappa \sigma^2 = \frac{\hat C_1 +7\hat C_2 +6\hat C_3 +\hat C_4}{\hat C_1+\hat C_2},}
this leads to larger deviations from the IdG in the charge conservation case.
These observations are reproduced in Appendix \ref{cons-formulas} within a simplified analytic model. 

Regarldess of the scenario for proton number fluctuations, one can see the overall signal becomes much weaker compared to that when considering fluctuations of all baryons.

\begin{figure*}[t]
    \includegraphics[width=0.9\textwidth]{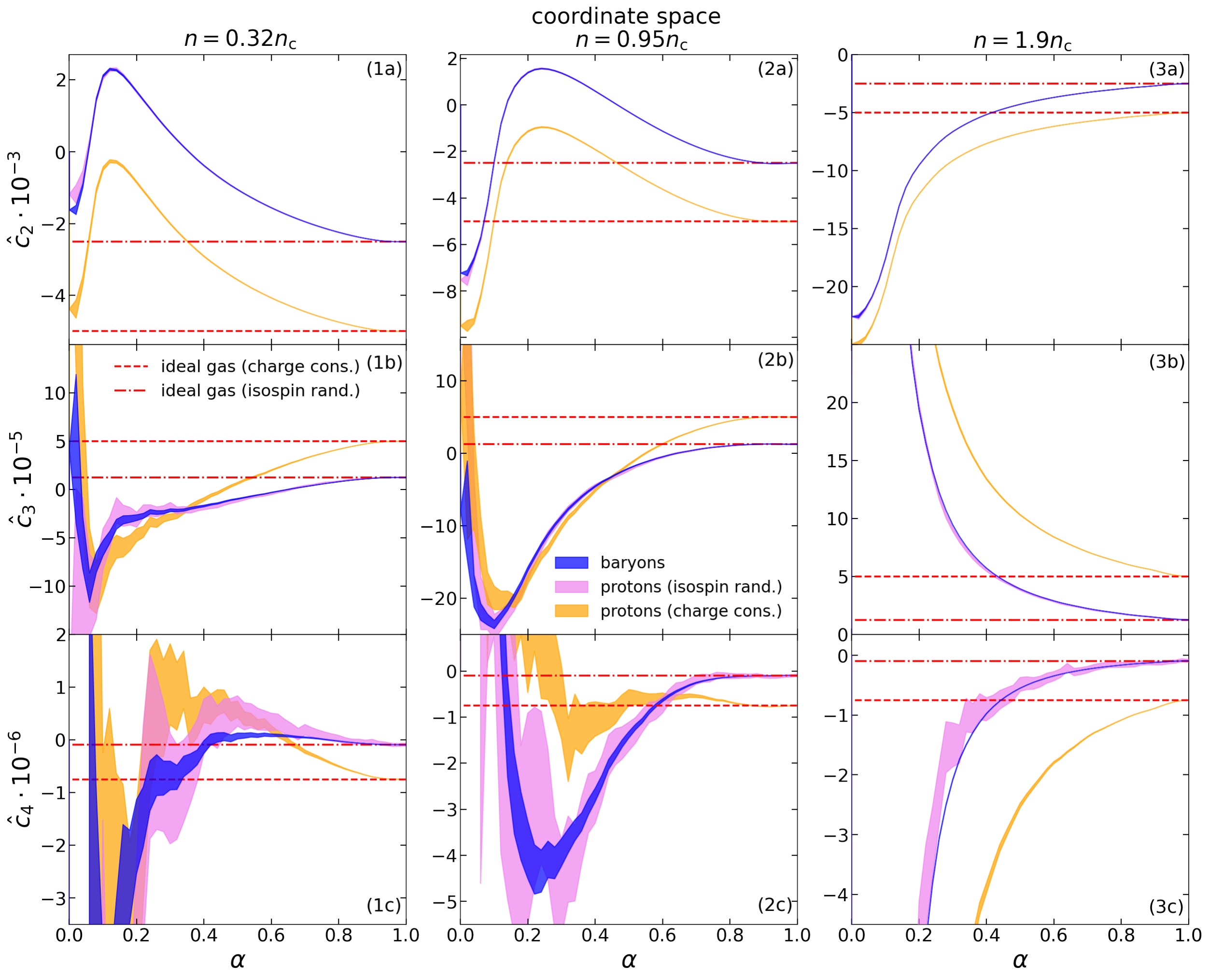}
    \caption{
    The same as in Fig. \ref{fig-factmoments} but for the {\it scaled} factorial cumulants \cite{Bzdak:2016sxg}.
    \label{fig-bzdakratios}
    }
\end{figure*}

\begin{figure*}[t]
    \includegraphics[width=0.95\textwidth]{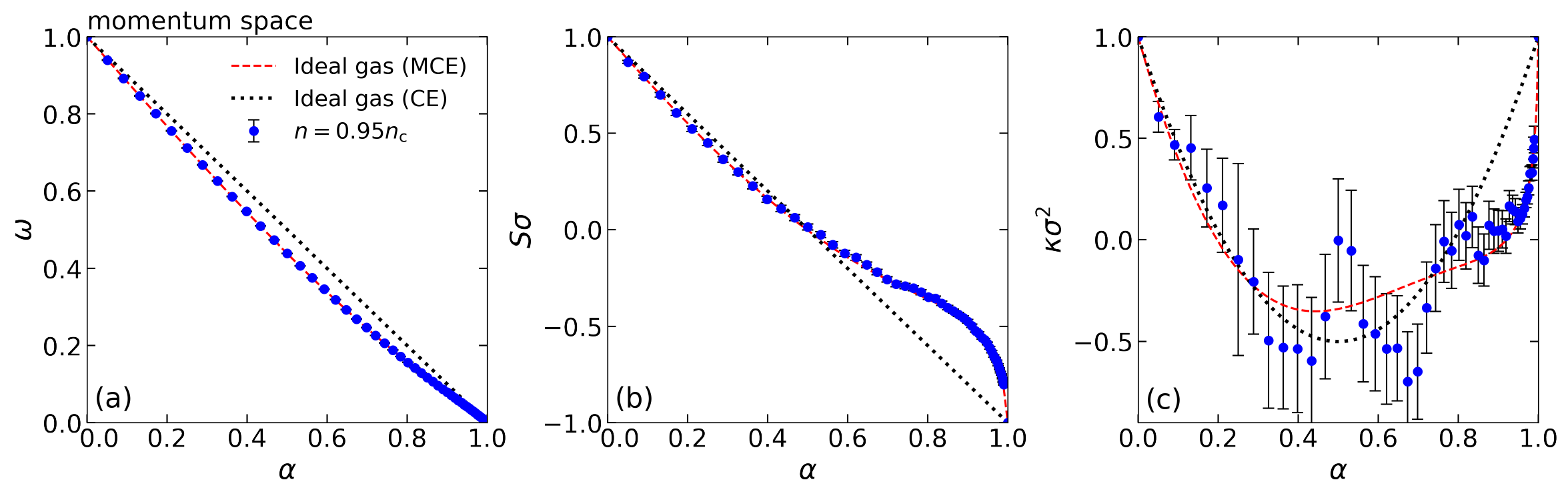}
    \caption{
    Acceptance fraction $\alpha$ dependence of scaled variance (a), skewness (b), and kurtosis (c) of particle number fluctuations in Lennard-Jones fluid within momentum space subsystem in the vicinity of the critical point ($T = 1.06T_{\rm c}$ and $n = 0.95n_{\rm c}$). The dashed red lines represent the micro-canonical ideal gas baseline. For scaled variance and skewness these lines are inside of the errors for MD simulation band.
    }
    \label{fig-momentumspace}
\end{figure*}

\subsection{Factorial Cumulants}

We now turn to factorial cumulants (\ref{eq:fc2})-(\ref{eq:fc4}), which have the advantage of removing self and all low-order correlations, thus probing multi-particle correlations. 
In this study, we analyze factorial cumulants for all particles (baryons), but also for protons in isospin randomization and charge conservation scenarios as described in Sec.~\ref{sec-conservation}. 

Figure \ref{fig-factmoments} presents $\alpha$-dependence of normalized factorial cumulants $\hat C_2/\mean{N}, \hat C_3/\mean{N}$, and $\hat C_4/\mean{N}$.
One sees that even at a relatively low density of $n = 0.32n_{\rm c}$ [panels (1a), (1b), (1c)] baryon factorial cumulants (blue curves) deviate significantly from the IdG baseline (lines).
$\hat{C}_2$ is enhanced and $\hat{C}_3$ is suppressed across all values of $\alpha$ in comparison with the IdG baseline, while $\hat{C}_4$ exhibits a non-monotonic $\alpha$-dependence dipping below the IdG level at low $\alpha$ and exceeding it at high $\alpha$. 
The signal is considerably weaker in proton cumulants.
In the isospin randomization scenario, one expects a suppression of correlations through a binomial factor, given by~\cite{Kitazawa:2012at}  
\eq{\label{eq:binom}
\hat{C}^{\rm p}_n=\frac{1}{2^n} \hat{C}^{\rm B}_n.
}
By comparing blue and magenta bands, one can see that this relation holds within the statistical errors, as expected.
Correspondingly, the magnitude of $\hat{C}_2 / \mean{N}$, $\hat{C}_3 / \mean{N}$, and $\hat{C}_4 / \mean{N}$ is suppressed by factors 2, 4, and 8, reducing the visibility of interaction (and CP) effects.
Proton cumulants in the charge conservation scenario show similar behavior as baryon cumulants, but with a reduced magnitude of factorial cumulants.

In the vicinity of the CP, $n = 0.95n_{\rm c}$, large deviations from the IdG baseline are observed [panels (2a), (2b), (2c)] in baryon number factorial cumulants. 
In particular, while $\hat C_2$ and $\hat C_3$ qualitatively resemble the behavior at $0.32n_{\rm c}$,
both of them show significantly larger deviations from the IdG baseline at near-critical density. 
The $\alpha$-dependence of the fourth order factorial cumulant, $\hat C_4$, is markedly different from that in the dilute regime, with $\hat C_4$ remaining well below the IdG line for all $\alpha \in (0.2,0.8)$ and reaching a pronounced minimum at $\alpha \approx 0.5$, where $\hat{C}_4 / \mean{N} \approx -11$. 
However, this signal is strongly diluted when proton factorial cumulants are considered instead, with the values of proton $\hat{C}_4/\mean{N}$ never reaching below $-2$.
In the charge conservation scenario, in particular, 
proton factorial cumulants shift closer to the IdG baseline compared to the full baryonic result, with $\hat{C}_4 / \mean{N}$ no longer showing a distinct signal despite the proximity to the CP.

Finally, at $n = 1.9n_{\rm c}$ [panels (3a), (3b), (3c)], correlations are dominated by repulsive interactions.
This leads to suppression of $\hat{C}_2 / \mean{N}$ and $\hat{C}_4 / \mean{N}$, and enhancement of $\hat{C}_3 / \mean{N}$ relative to the IdG baseline.
This result is consistent with previous calculations incorporating baryon repulsion via the excluded volume effect~\cite{Vovchenko:2021kxx}.
Notably, at high density, the $\alpha$-dependence of all three factorial cumulants is monotonic for both baryon and proton number distributions.

\subsection{Scaled factorial cumulants}

In addition to the ordinary normalized factorial cumulants, $\hat{C}_n / \mean{N}$ discussed above, a different normalization has been suggested in \cite{Bzdak:2016sxg}, where the so-called couplings 
\eq{
\hat{c}_n \equiv \hat{C}_n/\langle N \rangle^n
} 
have been introduced.
In this case, the factorial cumulants are normalized by the corresponding numbers of particle tuples.
A useful feature of the couplings is that studying their acceptance dependence allows one to distinguish short- and long-range correlations in the system.
Namely, in the absence of short-range correlations in the system, the couplings are independent of the chosen acceptance~\cite{Bzdak:2025rhp}.
In particular, this scaling holds if all correlations are driven by any combination of global baryon conservation, volume fluctuations, or efficiency.
Thus, the deviations from this scaling would indicate the presence of nontrivial short-range correlations, such as those due to the presence of the CP.

Figure~\ref{fig-bzdakratios} shows the coordinate space dependence of scaled factorial cumulants $\hat{c}_2, \hat{c}_3$, and $\hat{c}_4$ for baryons and protons, where they are depicted as a function of $\alpha$.
As before, the results are shown for the three representative particle number densities $n$. 
In the ideal gas case, where the only source of correlations is global baryon (or charge) conservation, the couplings are independent of $\alpha$ and shown by the horizontal lines.
Their values are determined solely by the total number of conserved particles, which is $N = 400$ for baryons, and $N = 200$ for protons with charge conservation.
The results from the LJ simulations show significant acceptance dependence for all densities considered.
This is a reflection of interactions, which are at a short (repulsion) and intermediate (attraction) range.
Focusing on the two-particle correlations (upper row in Fig.~\ref{fig-bzdakratios}), one can see that two-particle correlations are enhanced at low density ($n = 0.32 n_c$) and suppressed at high density ($n = 1.9 n_c$), reflecting the dominance of attractive and repulsive interactions, respectively.
Near the CP ($n = 0.95 n_c$), there is an interplay of attractive and repulsive interactions, where attraction and repulsion lead to a depletion~($\alpha \lesssim 0.1$) and enhancement~($\alpha \gtrsim 0.1$) of two-particle correlations, respectively.
Similar behavior is observed for third- and fourth-order couplings, although here the structure is a bit more complicated due to large statistical errors.

One can see that the couplings of baryons and of protons under isospin randomization coincide within errors.
This is an expected result.
As discussed before, factorial cumulants of protons are related to those of baryons through binomial folding~[Eq.~\eqref{eq:binom}].
When binomial folding is performed, it leaves the couplings unchanged~\cite{Bzdak:2025rhp}, and this is what we observe in our MD simulations.

In the case of conserved proton number (charge conservation), the numerical values of the couplings are different from baryons, mainly due to different total numbers of baryons and charges, meaning that the magnitude of the baryon and charge conservation effect is different.
Nevertheless, the proton couplings show the same qualitative dependence on the acceptance in both the isospin randomization and charge conservation scenarios, indicating that proton couplings are a robust probe of correlations among baryons.

To summarize, we observe that the presence of interactions leads to a clear violation of the scaling $\hat{c}_n = \rm const$ with acceptance.
In addition, the analysis of two-particle correlations of protons can establish whether attractive or repulsive interactions play the stronger role in a particular system. 
Although the analysis of couplings may not necessarily be sufficient to pinpoint the location of the CP~(e.g. the results for $\hat{c}_2$ for $n = 0.32n_c$ and $n = 0.95n_c$ are qualitatively similar), it seems that observing a non-flat dependence of $\hat{c}_n$ could give a strong indication for CP's existence.
It should be noted that the above analysis is performed in coordinate space, whereas the experiment measures fluctuations in the momentum space.
In the following, we thus also study the momentum-space acceptance.

\begin{figure*}[t]
    \includegraphics[width=0.9\textwidth]{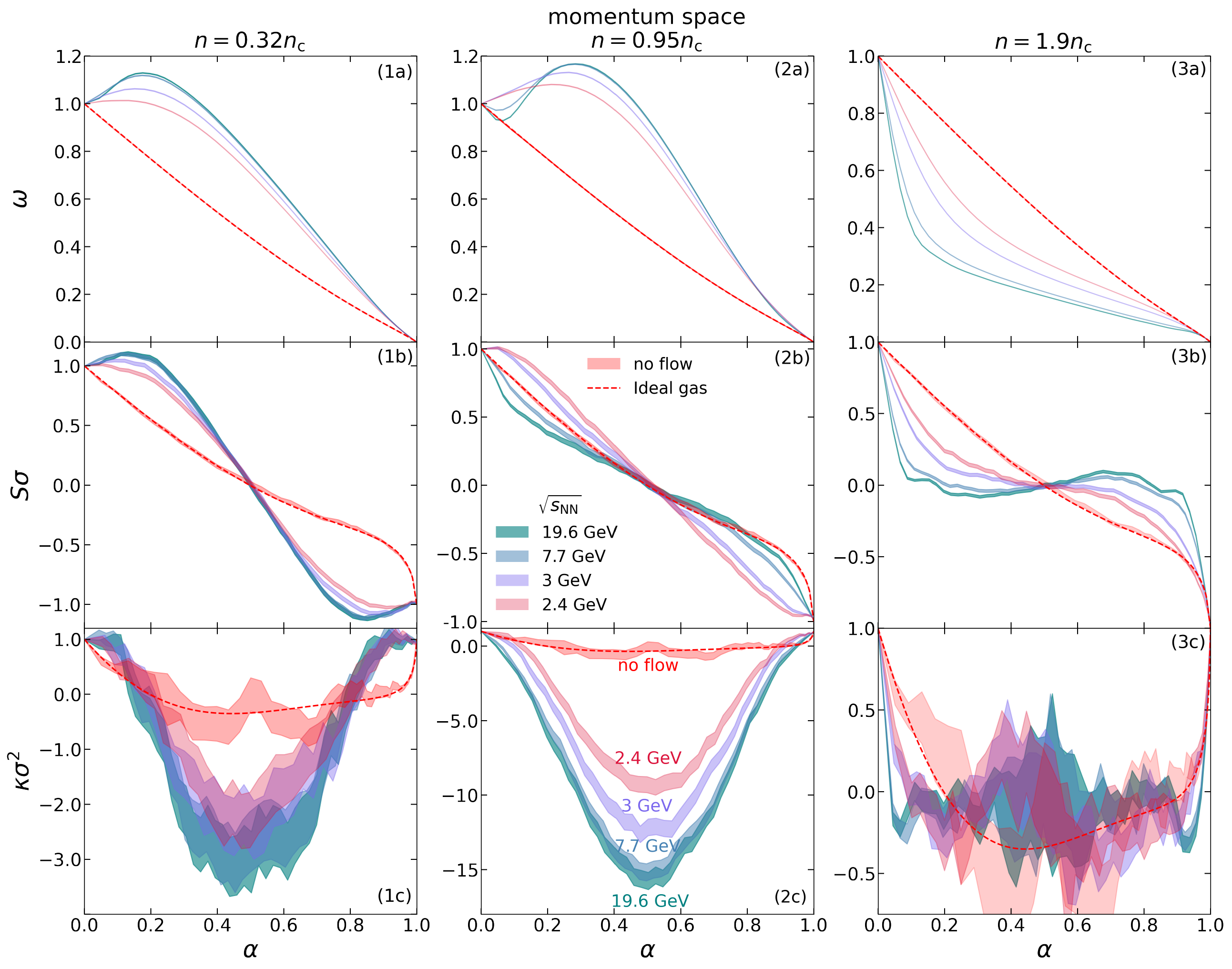}
    \caption{Acceptance fraction $\alpha$ dependence of scaled variance (top), skewness (middle), and kurtosis (bottom) of particle number fluctuations in Lennard-Jones fluid within momentum space subsystem at $T=1.06 T_c$ and $n/n_c=0.32, 0.95, 1.9$ (left to right).
    Colored bands show results at several collision energies $\sqrt{s_{\rm NN}}=2.4, 3.0, 7.7, 19.6$~GeV
    including collective flow as specified in Eq. (\ref{eq:ytot}). The red dashed curves denote the microcanonical ideal-gas baseline; the shaded red region indicates the no-flow case.}
    \label{fig-moments-snn-alpha}
\end{figure*}

\section{Fluctuations in Momentum Subspace}
\label{sec:Results_MS}

\subsection{Momentum Space Cumulants}
\label{sec:Results_MS_A}
We now turn to the study of fluctuations in momentum subspace, which is more relevant for relativistic heavy-ion collision experiments. 

First, we discuss the results without flow and the ideal-gas baseline expectations.
We note that the ideal-gas baseline for fluctuations in momentum space is somewhat more involved than the binomial distribution we worked with for coordinate space.
The reason is that our LJ simulations are performed in the microcanonical ensemble (MCE), which entails exact conservation of the total energy and the three components of momentum in addition to particle number conservation, and it affects the cumulants in momentum acceptance.
In~\cite{Kuznietsov:2022pcn}, we derived the expressions for the variance of particle number in longitudinal momentum acceptance, which takes into account exact conservation of energy and momentum.
Utilizing the same technique, here we derive the corresponding expression also for third and fourth-order cumulants~(the details can be found in Appendix~\ref{momentumIdG}).
For a large total particle number $N$, one can derive the following expression for the momentum space cumulant ratios $R_{nm} \equiv \kappa_n / \kappa_m$ within the ideal gas in MCE:
\eq{\label{eq:mce}
R^{\rm id}_{nm, \rm MCE}(\alpha) = R^{\rm id}_{nm, \rm CE}(\alpha) + \Delta R_{nm}^{\rm id}(\alpha).
}
Here $R^{\rm id}_{nm,\rm MCE}(\alpha)$ is the IdG baseline for a given cumulant ratio within the MCE, and $R^{\rm id}_{nm,\rm CE}(\alpha)$ is the same quantity within the CE, where it coincides with cumulant ratios from the binomial distribution.
The last term $\Delta R_{nm}^{\rm id}(\alpha)$ is the microcanonical correction due to energy and momentum conservation, see Appendix~\ref{momentumIdG} for explicit formulas.

Figure~\ref{fig-momentumspace} depicts the values of scaled variance, skewness, and kurtosis of baryon number fluctuations in momentum space subsystem evaluated in the IdG within the MCE~(dashed red lines) and CE~(dotted black lines), and compared to the LJ fluid simulations near the CP~($n = 0.95 n_c$).
The momentum space cut is performed in the longitudinal velocity, $|v_z| < v_z^{\rm cut}$, and the results are presented as a function of acceptance fraction $\alpha$, defined here as $\alpha = \mean{N}_{v_z^{\rm cut}} / N$.
Comparing CE and MCE results, one can see that energy–momentum conservation introduces subtle modifications to the behavior of the cumulants, which become larger at $\alpha > 0.5$, where energy conservation effects become stronger.
Comparison with the LJ fluid simulations shows excellent agreement of the MCE IdG baseline and numerical simulations.
This indicates that no effects of the CP~(or of any interaction at all) are present in momentum space cumulants, at least up to fourth order.
This generalizes our earlier results obtained on the level of second-order cumulants.\footnote{In our first work \cite{Kuznietsov:2022pcn}, momentum-space cumulants were computed through time averaging (instead of ensemble averaging) and a small dependence on the density was observed.}

\subsection{Cumulants with Collective Flow}

As shown in the previous subsection, momentum-space cumulants exhibit no clear interaction or CP signal in a purely equilibrium, box setup scenario.
The reason is that correlations due to interactions exist in coordinate space, and in a box setup there is no correlation between particles' momenta and coordinates, i.e. particle momentum distributions are the same at each coordinate inside the box.
However, in heavy-ion collisions, the presence of collective flow is well established, and the flow does correlate the coordinates and momenta.
In our previous work, we developed a simplified model to incorporate longitudinal flow and studied the resulting fluctuations in rapidity acceptances.
We found that the presence of flow allows one to recover critical point signals in the scaled variance inside rapidity acceptances depending on the collision energy, at least in this simplified model of flow.
Here we use the same model of flow to study the behavior of third- and fourth-order cumulants.

Following Ref.~\cite{Kuznietsov:2024xyn}, we introduce the longitudinal flow by adding a collective velocity component to each particle, which, as motivated by the Bjorken flow picture, is proportional to its longitudinal coordinate. 
The resulting rapidity of a particle is a sum of thermal and collective components:
\eq{\label{eq:ytot}
y = \frac{2 y_{\rm cm}^{\rm beam}}{\tilde L} \, \tilde z^{\rm LJ} + \sqrt{\frac{T_{\rm frz}}{m_N \tilde T}} \, \tilde v_z^{\rm LJ}.
}
Here $y$ is the final rapidity of a given particle from a given event, $\tilde z^{\rm LJ}$ and
$\tilde v_z^{\rm LJ}$ are, respectively, the reduced coordinate and (thermal) velocity of this particle along the longitudinal direction from the simulation,
and $y_{\rm cm}^{\rm beam}$
is the beam rapidity in the center-of-mass frame of the collision given by 
\eq{
y_{\rm cm}^{\rm beam}(\sNN)={\rm ln}\left[\frac{\sqrt{s_{\rm NN}}+\sqrt{s_{\rm NN}-4m_N^2}}{2 m_N}\right].
}
The temperature $T_{\rm frz}$ would roughly correspond to the thermal freeze-out temperature. 
In this work, we use a constant $T_{\rm frz} = 150$~MeV throughout for simplicity. 
We checked that the results are not very sensitive to the choice of $T_{\rm frz}$ value.
Finally, $m_N = 938$~MeV/$c^2$ is the nucleon mass.

We note that this is a highly simplified modeling of collective flow, and we do not attempt to make quantitative comparisons with data, but rather analyze the effects of collective flow under the most optimal conditions possible.

\subsubsection{$\alpha$ dependence}

We first analyze the $\alpha$-dependence of momentum-space cumulants in the presence of collective flow at different collision energies.
In practice, we apply various rapidity cuts, $|y| < y_{\rm cut}$ to calculate the number $N_{\rm acc}$ of accepted particles, and calculate the value of $\alpha$ as the fraction of the mean number of accepted particles relative to the total number of particles, $\alpha = \mean{N}_{\rm acc} / N$.
The expectation is that momentum-space results will approach coordinate-space ones for sufficiently strong collective flow, i.e., at high collision energies.

Figure~\ref{fig-moments-snn-alpha} shows the uncorrected scaled variance $\omega$, skewness $S\sigma$, and kurtosis $\kappa\sigma^{2}$ as functions of the acceptance fraction $\alpha$ for four collision energies $\sqrt{s_{\rm{NN}}}=2.4,3.0,7.7,19.6$ GeV.
In the absence of flow (red bands), all fluctuation measures coincide with the ideal-gas baseline (Sec.~\ref{sec:Results_MS_A}) within statistical errors. As $\alpha\to0$ and $\alpha\to1$, the cumulants approach, respectively, the Poisson limit ($\omega=S\sigma=\kappa\sigma^{2}=1$) and the full-acceptance limit ($\omega=0, S\sigma=-1, \kappa\sigma^{2}=1$).
At intermediate $\alpha$, flow induces pronounced deviations from the IdG baseline that grow with $\sqrt{s_{\rm{NN}}}$, signaling non-Gaussian behavior. At sufficiently high $\sqrt{s_{\rm{NN}}}$ the $\alpha$-dependence becomes nearly antisymmetric about $\alpha=0.5$ for $S\sigma$ and symmetric for $\kappa\sigma^{2}$, mirroring the structure expected for coordinate-space acceptance where energy-momentum conservation effects are negligible even at $\alpha\gg 0.5$.
Results at higher collision energies are very similar to $\sNN=19.6$ GeV and are not shown for brevity.

\subsubsection{Cumulants at fixed $y_{\rm cut}$}

\begin{figure*}[t]
    \includegraphics[width=0.95\textwidth]{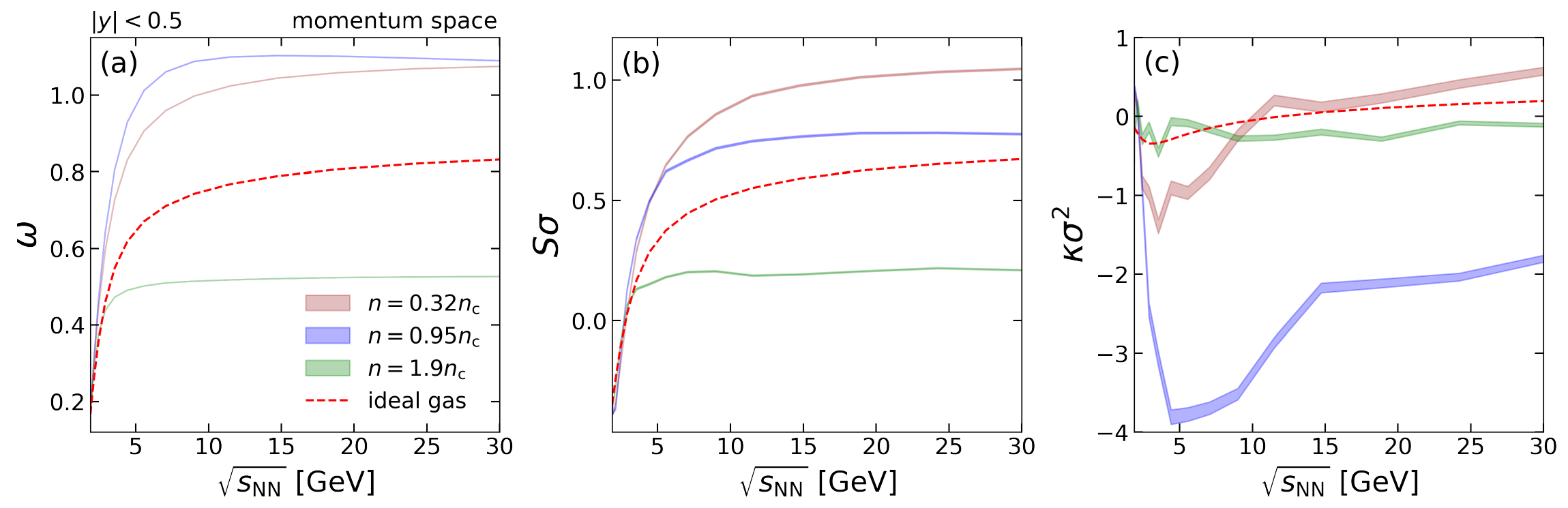}
    \caption{Scaled varience, skewness and kurtosis in momentum space with collective flow
    at a fixed rapidity cut of $y_{\rm cut} = 0.5$ which implies an acceptance fraction $\alpha$ which varies with $\sNN$.
    }
    \label{fig-moments-snn}
\end{figure*}

Here we study fluctuations at fixed rapidity cut $y_{\rm cut}$ as opposed to fixed $\alpha$.
This is a more natural setting for experimental analysis, where a fixed rapidity cut is used and implies that the acceptance fraction $\alpha$ is a function of $\sNN$.
Namely, for a fixed $y_{\rm cut}$, the acceptance fraction $\alpha \equiv \mean{N}/N$ is a monotonically decreasing function of $\sqrt{s_{\rm NN}}$, since fewer particles populate the fixed midrapidity window at higher energies.
This induces a non-monotonic interplay between collective flow and finite-size (acceptance) effects.

The three fluctuation measures are shown in Fig.~\ref{fig-moments-snn} as functions of $\sqrt{s_{\rm NN}}$ for $y_{\rm cut}=0.5$.
The ideal gas baseline taking into account baryon number and energy-momentum conservation effects~[Eq.~(\ref{eq:mce})] is presented by the dashed red curves (the baseline is density independent).
The lowest possible collision energy, $\sqrt{s_{\rm NN}}=2m_N$, corresponds to the absence of collective flow in our model, where $y_{\rm cut} = 0.5$ yields a relatively large acceptance fraction, $\alpha\simeq 0.86$.
As expected, at $\sqrt{s_{\rm NN}}\approx 2m_N$ all three measures for all densities coincide with their ideal-gas values at $\alpha\simeq 0.86$.
With increasing energy, $\omega$ and $S\sigma$ increase rapidly while $\kappa\sigma^2$ shows a non-monotonic behavior for a given density. 
This is a result the weakening of global-conservation suppression as $\alpha$ decreases with $\sqrt{s_{\rm NN}}$ (at fixed $y_{\rm cut}$).
In the presence of collective flow, all three observables deviate from the ideal-gas baseline.
These deviations are already sizable at $\sqrt{s_{\rm NN}}\sim 3$--$4~\text{GeV}$ and are strongest in the range $\sqrt{s_{\rm NN}}\sim 5$--$11~\text{GeV}$ (the precise location depends on the observable) for the considered $y_{\rm cut}=0.5$.

Notably, the scaled variance and the skewness do not show a sizable signal of the CP ($n = 0.95n_{\rm c}$) as compared to the calculation at a lower density ($n = 0.32n_{\rm c}$), see Fig.~\ref{fig-moments-snn}(a,b).
A strong signal of criticality is seen in the kurtosis, however, see Fig.~\ref{fig-moments-snn}(c).
At near-critical density $n = 0.95n_{\rm c}$, kurtosis reaches a large negative value $\kappa \sigma^2 \approx -4$ in energy range $\sqrt{s_{\rm NN}} \simeq 5-10$ GeV substantially below $\kappa \sigma^2 \approx -1.1$ at $n = 0.32n_{\rm c}$ and $\kappa \sigma^2 \approx -0.3$ at $n = 1.9n_{\rm c}$ in the same energy range. 
Observing a large negative kurtosis at $\sqrt{s_{\rm NN}} \sim 5$--$10$ GeV of baryon-number fluctuations can thus constitute a clear signal of criticality.

\begin{figure*}
\includegraphics[width=0.9\textwidth]{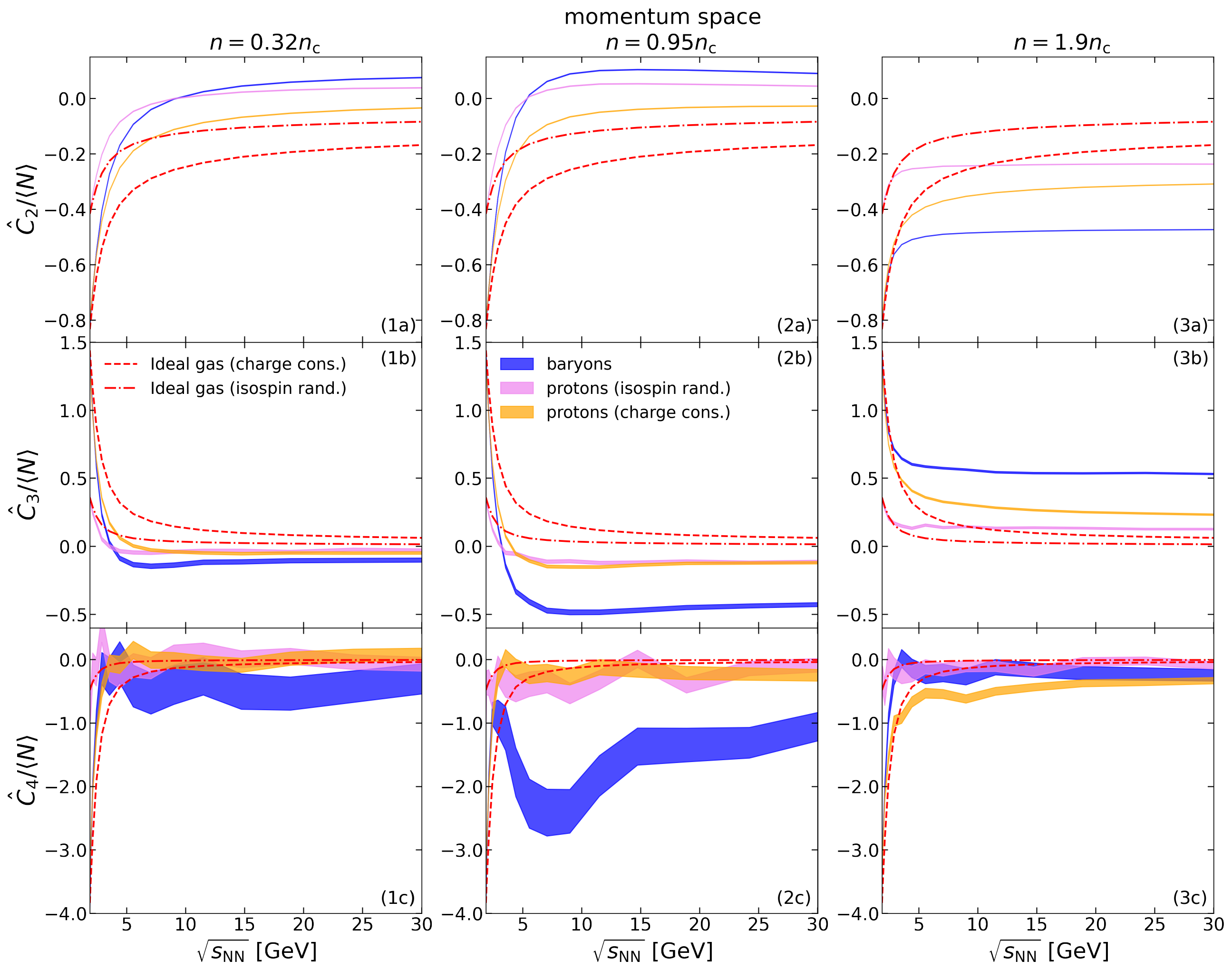}
    \caption{Normalized factorial cumulants $\hat{C}_2/\langle N\rangle$ (top), $\hat{C}_3/\langle N\rangle$ (middle), and $\hat{C}_4/\langle N\rangle$ (bottom) in the Lennard–Jones fluid at $T = 1.06T_{\rm c}$ and $n/n_c=0.32, 0.95, 1.9$ (left to right), evaluated within a momentum-space subsystem at $\tilde t = 50$ as functions of collision energy $\sqrt{s_{\rm NN}}$ with collective flow included according to Eq.~(\ref{eq:ytot}). Baryon-number fluctuations are shown as blue bands, while proton-number fluctuations with isospin randomization and with additional charge conservation are represented by pink and yellow bands, respectively. The corresponding ideal-gas baselines are indicated by dashed (isospin randomization) and dash-dotted (charge conservation) lines.
    \label{fig-factorial-snn}}
\end{figure*}

\subsubsection{Factorial cumulants at fixed $y_{\rm cut}$}

Figure~\ref{fig-factorial-snn} presents normalized factorial cumulants as functions of $\sqrt{s_{\rm NN}}$ for the three considered densities $n/n_c=0.32,0.95,1.9$ in the presence of collective flow.
As in Fig.~\ref{fig-moments-snn} we use a fixed rapidity window $y_{\rm cut}=0.5$.
Here we compare the factorial cumulants of (i) baryons, with those of (ii) protons with isospin randomization and (iii) protons with global charge conservation.

In the vicinity of the CP, at $n=0.95n_{\rm c}$ 
[panels (2a)-(2c)]
full baryon factorial cumulants (blue bands) as a function of collision energy show a strong signal of CP at intermediate energies where flow-induced coordinate-momentum correlations are strong and $\alpha$ is still moderate. In particular, large deviations from the ideal gas baseline, as collision energy is increased, emerge at around $\sqrt{s_{\rm NN}} \simeq 7-10$ GeV, namely $\hat C_2$ is enhanced, $\hat C_2/\mean{N}\approx 0.5$ is positive despite baryon conservation, while $\hat C_3$ and $\hat C_4$ are notably negative, e.g. $\hat C_3/\mean{N} \approx - 0.5$ and $\hat C_4/\mean{N} \approx -2.0$-$2.6$. This mirrors the behavior of the corresponding standard cumulants in Fig.~\ref{fig-moments-snn} and reflects enhanced fluctuations near the CP once flow correlates coordinate and momentum subspaces.

In dilute ($0.32n_c$) and dense ($1.9n_c$) systems, away from the CP, the deviations from the IdG baseline are weaker. In the dilute system, modest interaction effects survive at intermediate energies, most visible in $\hat C_2/\mean{N}$.
In the dense system, interaction effects are strong and all three factorial cumulants are systematically offset from IdG with $\sqrt{s_{\rm NN}}$-dependence being largely monotonic.
With increasing $\sqrt{s_{\rm NN}}$ (and thus decreasing $\alpha$) the curves approach the IdG expectation.

Switching from baryons to protons weakens the CP signal in the factorial cumulants. 
In the isospin randomization the proton and baryon factorial cumulants are expected to be related by the binomial acceptance factor $2^{k-1}$~[Eq.~(\ref{eq:binom})].
We checked that this relation holds in our simulations within error bars.
Accordingly, the CP is weakened in the isospin randomization case: instead of $\hat C_4/\mean{N} \approx -2.0$-$2.6$ for baryons, one observes $\hat C_4/\mean{N} \approx -0.5$-$0.65$ for protons.

The conserved proton number scenario (yellow bands) coincides with the full baryon picture and the ideal-gas line at small energies. At $\sNN>4$ GeV it deviates from it more than in the isospin randomization case, but is suppressed compared to the full system. 

\subsubsection{Acceptance dependence of scaled factorial cumulants}

\begin{figure*}[t]
\includegraphics[width=0.9\textwidth]{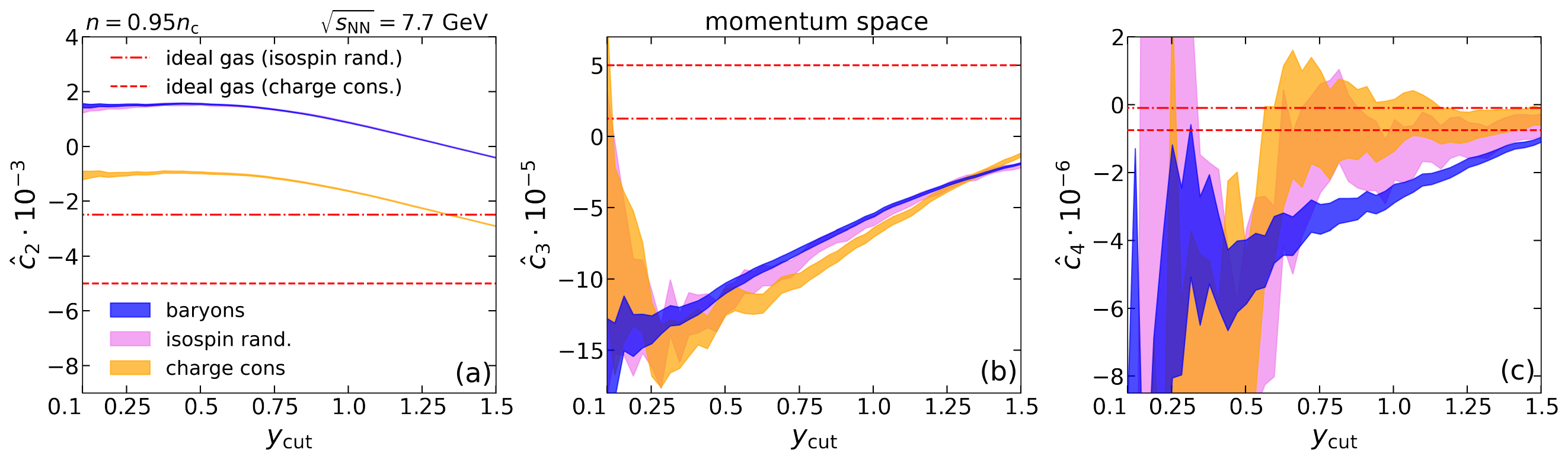}
\caption{Scaled factorial cumulants near the CP density as a function of rapidity cut at $\sNN = 7.7$ GeV in momentum space with collective flow. The choice $y_{\rm cut} \leq 1.5$ is motivated by the fact that large acceptances are not available in experiments. Note that large uncertainties at small $y_{\rm cut}$ arise from division by powers of $\mean{N} \to 0$.
}
    \label{fig-fact-s77}
\end{figure*}

Here we analyze the dependence of scaled factorial cumulants (couplings) $\hat{c}_n$ 
on the rapidity cut $y_{\rm cut}$ at the energy $\sNN = 7.7$ GeV.
This energy is the lowest one from the RHIC-BES program in collider mode.
We focus our analysis here on the near-critical density $n=0.95n_{\rm c}$.
% since critical effects should be strongest there.
The results for $\hat{c}_2$, $\hat{c}_3$, and $\hat{c}_4$ are shown in panels (a), (b), and (c) of Fig.~\ref{fig-fact-s77}.

Panel (a) shows almost flat behavior of $\hat{c}_2$ for $y_{\rm cut} < 0.75$, both for baryons and for protons. 
The results for baryon and protons under isospin randomization are consistent with each other, as expected from Eq.~(\ref{eq:binom}).
The results for protons under charge conservation are shifted relative to baryons by approximately constant factor.
This is consistent with the result derived in Appendix \ref{cons-formulas} within SAM approach.
At larger $y_{\rm cut}$ the lines approach IdG baseline (valid for full acceptance) monotonically, as more and more particles are captured by the momentum cut.
We note that the experimentally relevant rapidity cut range is $y_{\rm cut} \lesssim 0.6$.
The analysis of RHIC-BES-I data~\cite{STAR:2021iop} in Ref.~\cite{Bzdak:2025rhp} similarly showed flat behavior of $\hat{c}_2$ for $y_{\rm cut} \lesssim 0.5$.

The third-order coupling $\hat{c}_3$ is shown in panel (b). 
Due to large uncertainties at small rapidity cuts, it is difficult to say something about how monotonic are these lines at $y_{\rm cut} < 0.25$. 
However, at $y_{\rm cut} > 0.25$ the lines show a clear non-flat, monotonic behavior, approaching the ideal gas limit in full acceptance from below.
The results there are consistent between baryon and both proton scenarios.
The non-flat behavior should be accessible experimentally at $y_{\rm cut} \lesssim 0.6$.
Due to large uncertainties it is not possible to verify this behavior with RHIC-BES-I data, but it should be accessible with RHIC-BES-II data.

Finally, panel (c) shows the fourth-order coupling $\hat{c}_4$.
We observe indications for a non-monotonic behavior of $\hat{c}_4$ for baryons at $0.25 < \alpha < 0.75$, although with sizable uncertainties.
The uncertainties for protons are larger and preclude any direct conclusions about their behavior.
However, protons under isospin randomization are expected to coincide with baryons.
Similarly to $\hat{c}_3$, the uncertainties at RHIC-BES-I are too large to draw any conclusions about the behavior of $\hat{c}_4$.
It is possible that at RHIC-BES-II the signal will be more accessible.

\section{Summary and outlook}
\label{sec:Summary}

We have performed a microscopic study of higher-order cumulants of particle number fluctuations near the critical endpoint of a first-order phase transition using molecular dynamics simulations of the classical Lennard-Jones fluid. 
We performed our simulations near the critical point at $n=0.95n_{\rm c}$, as well as in dilute ($n=0.32n_{\rm c}$) and dense ($n=1.9n_{\rm c}$) regimes.
We analyzed ordinary and factorial cumulants, differences between proton and baryon systems, as well as coordinate and momentum space acceptances.
Our simulations were done for a system of 400 particles, reflecting the maximum number of participant nucleons in heavy-ion collisions.
The results show that large effects of the critical point on the crossover side are visible in higher-order cumulants, especially the kurtosis, but with many caveats.
Among the many results obtained, one can highlight the following:

\begin{itemize}
    \item Equilibration. Cumulants of different order show similar finite-time effects. In particular, we observe that they equilibrate on comparable time scales even close to the critical point ($n=0.95n_{\rm c}$), as visible in Fig.~\ref{fig-timedep}. Our results thus indicate that critical slowing down does not significantly inhibit equilibration of higher-order cumulants relative to the second-order cumulants.
    \item Finite-size effects. The cumulants show strong finite-size effects near the CP, with higher-order cumulants being more sensitive to the system size than the second-order cumulants. This is particularly evident when comparing the values extracted from MD simulations with the grand-canonical susceptibilities (Fig.~\ref{fig-virial-cumulants}).
    \item Despite the strong finite-size effects, a clear signal of the crossover region is observed in the kurtosis, where significantly negative values of $\kappa \sigma^2 \simeq -10$ to $-15$ are observed in simulations at $n=0.95n_{\rm c}$, as shown in Fig.~\ref{fig-prot-standcumulants}. These large negative values are also reflected in fourth-order factorial cumulants (Fig.~\ref{fig-factmoments}).
    \item Signals of criticality in ratios of ordinary cumulants are diluted when protons are considered instead of baryons. In the isospin randomization case in particular, the selection of protons instead of baryons acts like an efficiency cut which leads to a Poissonization of the cumulants, especially pronounced in higher-order cumulants.
    \item Factorial cumulants disentangle effects of multi-particle correlations of different order and show the expected critical point behavior more clearly than the ratios of ordinary cumulants. Using scaled factorial cumulants $\hat{c}_n$ reveals very similar behavior for both baryons and protons and may be more suitable for experimental analysis.
    \item The CP signal disappears in momentum space acceptance without collective flow in all cumulants. However, implementing a Bjorken-like collective flow model restores a significant part of the CP signal in momentum space. The behavior of the CP signal in momentum space with collective flow is qualitatively similar to the coordinate-space acceptance.
\end{itemize}

We also discuss our results in the context of recent experimental data from RHIC-BES-II~\cite{STAR:2025zdq}.
The bare data presented in Ref.~\cite{STAR:2025zdq} showed limited evidence for CP in ordinary cumulants.
However, interesting features have been observed in factorial cumulants, especially compared to non-critical baselines.
The second-order factorial cumulant $\hat{C}_2/\hat{C}_1$ is negative throughout the whole collider energy range.
At $\sNN \gtrsim 10$~GeV, the result is consistent with baryon conservation effects supplemented by repulsive baryon interactions. Qualitatively, this corresponds to our simulations in the dense system ($n=1.9n_{\rm c}$), where repulsive interactions dominate (see panel (3) in Fig.~\ref{fig-factorial-snn}).
However, a clear change of trend is visible at $\sNN \lesssim 10$~GeV, with $\hat{C}_2/\hat{C}_1$ starting to increase as collision energy is reduced.
Qualitatively, this is consistent with our simulations at smaller densities, both at $n=0.32n_{\rm c}$ and $n=0.95n_{\rm c}$, indicating a shift from repulsive to attractive interactions.
Similarly, the data for $\hat{C}_3/\hat{C}_1$ show a non-monotonic behavior with a peak at $\sNN \sim 10$~GeV.
The right side of the peak is described by a non-critical baseline with repulsion, while the left side indicates a negative deviation from the baseline.
This is again consistent with a shift from $n = 1.9n_{\rm c}$ simulations (repulsion-dominated) to $n = 0.95n_{\rm c}$ (interplay of repulsion and attraction) and to $n = 0.32n_{\rm c}$ (attraction-dominated).

The data for $\hat{C}_4 / \hat{C}_1$ are largely consistent with zero although with sizable error bars, constraining the values to $-0.25 < \hat{C}_4 / \hat{C}_1 < 0.5$. 
On the first glance, it may be interpreted as evidence against the critical point. 
However, it is important to note that measurements are performed for protons, where, as our simulations show, the signal is significantly suppressed compared to baryons.
As seen in Fig.~\ref{fig-factorial-snn}, while for baryons one can observe $\hat{C}_4 / \hat{C}_1 \sim -1$-$2$ at $\sNN \sim 3$-$10$~GeV, the same signal for protons can be much smaller, with $\hat{C}_4^p / \hat{C}_1^p \sim -0.25$-$1$ depending on the treatment of electric charge conservation effects.
The observed values by STAR can thus be consistent with critical point effects even under the relatively simplified treatment presented in our work. 

We argue that further information can be obtained by analyzing the scaled factorial cumulants $\hat{c}_n$ as a function of rapidity acceptance. 
Our simulations predict consistent behavior of $\hat{c}_n$ between baryons and protons.
Furthermore, as argued in Ref.~\cite{Bzdak:2025rhp}, non-flat behavior of $\hat{c}_n$ elucidates the effects of local correlation due interactions (such as those due to the CP) and absorbs the effects of volume fluctuations through a trivial constant shift of the curve.
Our simulations predict a flat behavior of $\hat{c}_2$ in experimentally reachable rapidity acceptance, while $\hat{c}_3$ and $\hat{c}_4$ show non-flat behavior~(Fig.~\ref{fig-bzdakratios}).
The analysis of RHIC-BES-II data can be used to test this prediction.

Our model here is too much simplified to be used for a quantitative analysis of experimental data.
Further work can be performed to bridge this gap.
For instance, one can utilize a more realistic collective flow prescription based on hydrodynamic simulations.
Simulations for different system sizes can be performed to study finite-size scaling of high-order cumulants.
Ultimately, the interactions responsible for the CP here can be incorporated into realistic transport models.

\begin{acknowledgments}

\emph{Acknowledgments.} 
We thank Greg Morrison and Anar Rustamov for fruitful discussions.
The authors acknowledge the use of the PhysGPU Cluster and the support from the Research Computing Data Core at the University of Houston to carry out the research presented here.
V.K. has been supported by the U.S. Department of Energy, 
Office of Science, Office of Nuclear Physics, under contract number 
DE-AC02-05CH11231.
V.A.K. and V.V. have been supported by the U.S. Department of Energy, 
Office of Science, Office of Nuclear Physics, Early Career Research Program under Award Number DE-SC0026065.
M.I.G. is thankful for support from the Simons Foundation.

\end{acknowledgments}

\appendix

\section{KN-Parametrized EoS}
\label{KolafaEOS}
A different approach to build an analytical, parametrized, dimensionless EoS for LJ matter was attempted in Ref.~\cite{KOLAFA19941}. It starts with the hard-sphere Carnahan–Starling equation of state in the form of the free energy
\eq{F_{\rm HS} = T\left[\frac{5}{3}{\rm ln}(1-\eta) + \frac{\eta(34-33\eta + 4\eta^2)}{6(1-\eta)^2}\right],}
where $\eta = (4\pi/3)\tilde n \tilde r^3$, with corresponding pressure
\eq{p_{\rm HS} = \tilde n \tilde T \left[\frac{1 + \eta + \eta^2 - (2\eta^3/3)(1+\eta)}{(1-\eta)^3}\right].}

The second virial coefficient $\tilde B_2(T)$ is analytically known \cite{Gottschalk:2019}. This gives us an opportunity to define a second-order correction to the HS free energy as
\eq{\tilde F_2 = \tilde F_{\rm HS} + \tilde n \tilde T (\tilde B_{2,LJ}(\tilde T) - \tilde B_{2,HS}(\tilde T))e^{-\gamma \tilde n^2},}
which yields a matching second virial coefficient with the LJ EoS and represents reasonable thermodynamics at large densities due to the phenomenological damping factor ${\rm exp}(-\gamma n^2)$.

However, to capture structure beyond the second virial coefficient, we can find a correction to $\tilde F_2$ in the form
\eq{\tilde F_{LJ} \approx \tilde F_2 + \Delta \tilde F = \tilde F_2 + \sum\limits_{i,j} C_{i,j} \tilde T^{i/2} \tilde n^j,}
where the correction $\Delta F$ is written as a power series in density and temperature with numerically determined coefficients from Ref.~\cite{KOLAFA19941}. The same reference gives the hard-sphere radius and damping constant $\gamma$.

Finally, using standard thermodynamic relations one can obtain the pressure and build susceptibilities as its derivatives.
\eq{\begin{split}
    \tilde p = \tilde p_{\rm HS} &+ \tilde n^2 T(1 - 2\gamma \tilde n^2){\rm exp}(-\gamma \tilde n^2)\Delta \tilde B_2 \\&+ \sum\limits_{i,j}j C_{i,j}\tilde T^{i/2} \tilde n^{j+1}.
\end{split}}

Knowing the EoS as a function of density gives one a possibility to express GCE susceptibilities using $\mu= \mu(\tilde T,\tilde n)$ and
\eq{\begin{split}
    \chi_n &= \tilde T^{n-4}\left(\frac{\partial^n \tilde p}{\partial \mu^n}\right)_{\tilde T,V} = \tilde T^{n-4}\frac{\partial^{n-1}}{\partial \mu^{n-1}}\left(\frac{\partial \tilde p}{\partial \tilde n}\frac{\partial \tilde n}{\partial \mu}\right)_{\tilde T,V} \\&=\tilde T^{n-4}\frac{\partial^{n-1}}{\partial \mu^{n-1}}\left(\frac{\partial \tilde p(\tilde T,\tilde n)}{\partial \tilde n}\chi_2(\tilde T,\tilde n)\right)_{\tilde T,V}
\end{split}}
where density derivatives can be obtained from the EoS and higher $\mu$ derivatives can be expressed through density derivatives to any order recursively in terms of lower-order susceptibilities.

\begin{widetext}
\section{Analytical Approximations for Proton Number Fluctuations in the Thermodynamic Limit}
\label{cons-formulas}
Two conservation scenarios discussed in our paper can be studied in connection with the full-system cumulants. The prescriptions are discussed in \cite{Savchuk:2019xfg} and \cite{Vovchenko:2020gne}.

The GCE proton susceptibilities are related to baryon susceptiblities by folding the latter with binomial distribution with probability $p = 1/2$. One obtains
\eq{\begin{split}
    \chi^p_1 = \chi_1^B/2,~~~\chi^p_2 = (\chi_2^B + \chi_1^B)/4,~~~\chi^p_3 = (3\chi^B_2 + \chi_3^B)/8.
\end{split}}
and central cumulants (both proton and baryon)
\eq{\kappa_1 = VT^3\alpha \chi_1,~~~ \kappa_2 = VT^3\alpha (1-\alpha) \chi_2,~~~\kappa_3 = VT^3\alpha(1-\alpha)(1-2\alpha) \chi_3}
one can find for second factorial cumulant
\eq{\left(\frac{\hat C^p_2}{\hat C^p_1}\right)_{\rm is.rand.} &=\frac{1}{2}\left(\frac{\hat C^B_2}{\hat C^B_1}\right) = \frac{\kappa^B_2-\kappa^B_1}{2\kappa^B_1} = -\frac{1}{2} + \frac{1-\alpha}{2}\frac{\chi_2^B}{\chi_1^B},
\\\left(\frac{\hat C^p_2}{\hat C^p_1}\right)_{\rm ch. cons.} &= \frac{\kappa^p_2-\kappa^p_1}{\kappa^p_1} = \frac{(1-\alpha)\chi^p_2 - \chi^p_1}{\chi^p_1} = -\frac{1+\alpha}{2} + \frac{1-\alpha}{2}\frac{\chi_2^B}{\chi_1^B},}
then the difference from the ideal-gas baseline will be equal for both cases
\eq{\left(\frac{\hat C^p_2}{\hat C^p_1}\right) - \left(\frac{\hat C^p_2}{\hat C^p_1}\right)_{\rm id} =\frac{\alpha}{2}+\left(\frac{\hat C^p_2}{\hat C^p_1}\right)_{\rm is. rand.} = \alpha + \left(\frac{\hat C^p_2}{\hat C^p_1}\right)_{\rm ch. cons.} = \frac{\omega + \alpha - 1}{2}}
which means that for the second factorial cumulant the deviation from the IdG baseline is the same.

However, this is not the case for the third factorial cumulant. Namely,
\eq{\left(\frac{\hat C^p_3}{\hat C^p_1}\right)_{\rm is. rand.} &= \frac{1}{4}\left(\frac{\hat C^B_3}{\hat C^B_1}\right) = \frac{\kappa^B_3-3\kappa^B_2+\kappa^B_1}{4\kappa^B_1} = \frac{1}{2} - \frac{3(1-\alpha)}{4}\frac{\chi_2^B}{\chi^B_1} +\frac{(1-\alpha)(1-2\alpha)}{4}\frac{\chi^B_3}{\chi^B_1}.
\\\begin{split}
\left(\frac{\hat C^p_3}{\hat C^p_1}\right)_{\rm ch. cons.} &= \frac{\kappa^p_3-3\kappa^p_2+\kappa^p_1}{\kappa^p_1}= \frac{(1-\alpha)(1-2\alpha)\chi^p_3 -3(1-\alpha)\chi^p_2 + \chi^p_1}{\chi^p_1} \\&= \frac{1+3\alpha}{2} -\frac{3(1-\alpha)(1+2\alpha)}{4}\frac{\chi^B_2}{\chi^B_1} + \frac{3(1-\alpha)(1-2\alpha)}{4}\frac{\chi^B_3}{\chi^B_1},
\end{split}}
the difference with the ideal-gas baseline will be
\eq{\left(\frac{\hat C^p_3}{\hat C^p_1}\right)_{\rm ch. cons.} - 2\alpha^2 -\left(\frac{\hat C^p_3}{\hat C^p_1}\right)_{\rm is.rand} + \frac{\alpha^2}{2}= \frac{3\alpha(1-\alpha)}{2}\left(1-\frac{\chi^B_2}{\chi^B_1}\right),}
which is larger than $0$ if $\chi^B_2/\chi^B_1 < 1$ or $\omega^B < 1 - \alpha$ and $\alpha \ne \{0,1\}$. It's a non-trivial result that at given acceptance difference between deviations from IdG in two conservation scenarios depends only on scaled variance value. 

It can be shown that factorial cumulants will have different deviations from the IdG baseline for $\hat C^p_4/\hat C^p_1$ and higher cumulants as well; however, these formulas will consist of mixed charge susceptibilities, which we are not discussing in our model.  

\section{Ideal Gas in the Microcanonical Ensemble}
\label{momentumIdG}

In coordinate space, energy conservation does not play any role if the particles are not interacting. Namely, for a given acceptance $0 \leq z/L \leq \alpha$ one will have
\eq{\omega^{\rm coord}_i(q_1,\dots,q_i) = \frac{1}{Z}\int\limits \prod^{3N}_{j=1} d p_j ~\delta\left(2mE - \sum p_j^2\right) \int \prod^{3N-i}_{l=1} d q_l = \frac{1}{L^i},}
\eq{F^{\rm coord, id}_i = \frac{N!}{(N-i)!} \int\limits_0^{\alpha L} d q_1 \dots  d q_i~ \omega^{\rm coord}_i(q_1,\dots,q_i) = \frac{\alpha^iN!}{(N-i)!} ,}
which corresponds to the factorial moments of an ideal gas with binomial distribution (see \cite{Potts1953-wn}).

In momentum space, following the method proposed in \cite{Kuznietsov:2022pcn}, one can compute the cumulants of an IdG while incorporating energy conservation. Specifically,
\eq{\label{A1}
F^{\rm id}_i(p_1,\dots,p_i) = \frac{N!}{(N-i)!} \int\limits_{-p_0}^{p_0} d p_1 \dots  d p_i~ \omega_i(p_1,\dots,p_i).
}

Since in the current work we consider cumulants up to fourth order, here we consider $i\leq 4$ particle probability distributions:

\eq{\begin{split}
&\omega_1(p_1) = \frac{R^{2-k}(R^2-p_1^2)^{\frac{k-3}{2}}\Gamma\left(\frac{k}{2}\right)}{\sqrt{\pi}\Gamma\left(\frac{k-1}{2}\right)},
\\&\omega_2(p_1,p_2) = \frac{R^{2-k}(R^2-p_1^2-p_2^2)^{\frac{k-4}{2}}(k-2)}{2\pi},
\\&\omega_3(p_1,p_2,p_3) = \frac{R^{2-k}(R^2-p_1^2-p_2^2-p_3^2)^{\frac{k-5}{2}}\Gamma\left(\frac{k}{2}\right)}{\pi^{3/2}\Gamma\left(\frac{k-3}{2}\right)},
\\&\omega_4(p_1,p_2,p_3,p_4) = \frac{R^{2-k}(R^2-p_1^2-p_2^2-p_3^2-p_4^2)^{\frac{k-6}{2}}(k-2)(k-4)}{4\pi^2}.
\end{split}}

Here, $R = \sqrt{2mE}$ and $k = 3N$. By using dimensionless coordinates $q = \frac{\sqrt{k}}{R}p$ and performing a variable transformation under the integral in \eqref{A1} with the appropriate Jacobian, one can obtain explicit expressions that are independent of $R$:

\eq{\begin{split}\label{A3}
&\omega_1(q_1) = \frac{k^{\frac{2-k}{2}}(k-q_1^2)^{\frac{k-3}{2}}\Gamma\left(\frac{k}{2}\right)}{\sqrt{\pi}\Gamma\left(\frac{k-1}{2}\right)},
\\&\omega_2(q_1,q_2) = \frac{k^{\frac{2-k}{2}}(k-q_1^2-q_2^2)^{\frac{k-4}{2}}(k-2)}{2\pi},
\\&\omega_3(q_1,q_2,q_3) = \frac{k^{\frac{2-k}{2}}(k-q_1^2-q_2^2-q_3^2)^{\frac{k-5}{2}}\Gamma\left(\frac{k}{2}\right)}{\pi^{3/2}\Gamma\left(\frac{k-3}{2}\right)},
\\&\omega_4(q_1,q_2,q_3,q_4) = \frac{k^{\frac{2-k}{2}}(k-q_1^2-q_2^2-q_3^2-q_4^2)^{\frac{k-6}{2}}(k-2)(k-4)}{4\pi^2}.
\end{split}}

The integrals in Eq.~\eqref{A1} can be computed analytically in the large-$N$ limit using Eq.~\eqref{A3} and expressed as functions of $\alpha = \mean{N}/N$. This method can be extended to compute cumulants of any order, incorporating microcanonical corrections that become significant for small systems and large acceptances. 

Since functions in \eqref{A3} in the infinite system limit gives $\omega_i({\sum q_i^2}) \sim \exp({\sum q_i^2}/2)$, which corresponds to the standard coordinate-space formulas for the IdG, one can use a formal expansion around $k \to \infty$ in the form:

\eq{
&\omega_i\left({\sum\limits^i_{j=1} q_j^2}\right)  = (2\pi)^{-i/2}\exp\left(\sum\limits^i_{j=1} q_j^2/2\right) + \sum \limits_{j = 1}^{\infty}~\frac{k^{-j}}{j!} \lim \limits_{k \to \infty} \frac{\partial^j \omega_i(k, \sum q^2)}{\partial~ (1/k)^j}. \label{A4}
}

Integration of \eqref{A4} over all $i$ coordinates leads to the convenient form:

\eq{\begin{split}
F^{\rm id}_{i, \rm MCE} &=F^{\rm id}_{i, \rm CE} +  N(N-1)\dots (N-i+1)\sum \limits_{j = 1}^{\infty} \frac{k^{-j}}{j!} \left[\prod\limits_{s=1}^i \int\limits^{q_{\rm cut}}_{-q_{\rm cut}} d q_s\right] \lim \limits_{k \to \infty} \frac{\partial^j \omega_i(k, \sum q^2)}{\partial~ (1/k)^j} \\&=
F^{\rm id}_{i,\rm CE} +  N(N-1)\dots (N-i+1)\sum \limits_{j = 1}^{\infty} \frac{k^{-j}}{j!} \lim \limits_{k \to \infty} \frac{\partial^j}{\partial~ (1/k)^j} \left[\prod\limits_{s=1}^i \int\limits^{q_{\rm cut}}_{-q_{\rm cut}} d q_s\right] \omega_i(k, \sum q^2)\\&=
F^{\rm id}_{i,\rm CE}+ k(k-3)\dots (k-3(i-1))\sum \limits_{j = 1}^{\infty} \frac{k^{-j}}{j!} \lim \limits_{k \to \infty} \frac{\partial^j}{\partial~ (1/k)^j}\frac{F^{\rm id}_i(k,q_{\rm cut})}{k(k-3)\dots (k-3(i-1))} %= C^{\rm id}_{i,\rm CE} + \Delta C^{\rm id}_i(k,q_{\rm cut}).
\label{A5}
\end{split}}
where $k=3N$ was used.

Since $\alpha = \mean{N}/N = \mathrm{erf}(q_{\rm cut}/\sqrt{2})$, we have $q_{\rm cut}(\alpha) = \sqrt{2} ~{\rm erf}^{-1}(\alpha)$, so one can express $F_i^{\rm id}(k,q_{\rm cut}) = F_i^{\rm id}(k,\alpha)$ without loss of generality. For large $k$, a truncated version of the series in \eqref{A5} can be used, namely
\eq{F^{\rm id}_{i, \rm MCE} = F^{\rm id}_{i,\rm CE} + \sum \limits_{r=1}^{i} 3^{i-r}s(i,r)\sum \limits_{j = 1}^{r} \frac{k^{r-j}}{j!} \lim \limits_{k \to \infty} \frac{\partial^j}{\partial~ (1/k)^j}\left[\frac{F^{\rm id}_i(k,q_{\rm cut})}{k(k-3)\dots (k-3(i-1))}\right] + O(1/k)\label{A6}}
where $s(i,r)$ is a Stirling number of the first kind.

In our case, we need $i\leq 4$, which can be found directly from Eq.~\eqref{A6}.
One can reconstruct the cumulants using the known relation between factorial moments and cumulants.
The cumulant ratios $R_{nm}^{\rm id} = \kappa_n^{\rm id} / \kappa_m^{\rm id}$ are independent of $k$ in the large-$k$ limit and can be similarly represented in terms of a microcanonical correction to the CE result:
\eq{
R^{\rm id}_{nm,\rm MCE}(\alpha) = R^{\rm id}_{nm,\rm CE}(\alpha) + \Delta R^{\rm id}_{nm}(\alpha),
}
namely.
Given that $\omega \equiv R_{21}$, $S\sigma \equiv R_{32}$, and $\kappa\sigma^{2} \equiv R_{42}$, one obtains
\eq{\begin{split}
    &\Delta \omega^{\rm id}(\alpha) = -\frac{2e^{-2{\rm erf}^{-1}(\alpha)^2}({\rm erf}^{-1}(\alpha))^2}{3\pi \alpha},
    \\&\Delta (S\sigma)^{\rm id}(\alpha) = - \frac{2e^{-{\rm erf}^{-1}(\alpha)^2}{\rm erf}^{-1}(\alpha)^2(6~e^{{\rm erf}^{-1}(\alpha)^2}\sqrt{\pi}~(2\alpha - 1)+{\rm erf}^{-1}(\alpha)-6~{\rm erf}^{-1}(\alpha)^3)}{3\sqrt{\pi}(3\pi~e^{2{\rm erf}^{-1}(\alpha)^2}(\alpha-1)\alpha+2~{\rm erf}^{-1}(\alpha)^2)}
    \\&\Delta (\kappa\sigma^{2})^{\rm id}(\alpha) = \frac{4~e^{-2{\rm erf}^{-1}(\alpha)^2}{\rm erf}^{-1}(\alpha)^2(27e^{2{\rm erf}^{-1}(\alpha)^2}\pi (1-5\alpha+5\alpha^2)+9~e^{{\rm erf}^{-1}(\alpha)^2}\sqrt{\pi}(2\alpha-1){\rm erf}^{-1}(\alpha)^2 + 36~{\rm erf}^{-1}(\alpha)^2)}{9\pi(3e^{2{\rm erf}^{-1}(\alpha)^2}\pi(\alpha -1)\alpha+ 2~{\rm erf}^{-1}(\alpha)^2)}
\end{split}}

The derived expressions provide a theoretical baseline, shown as the red dashed line in Fig.~\ref{fig-momentumspace}, which matches MD results without collective flow. This agreement confirms the absence of interactions and validates our approach. Since these results are derived purely from statistical mechanics, they offer a crucial reference for distinguishing critical fluctuations from conservation effects.

\end{widetext}

\bibliography{main}
\end{document}